\documentclass{aa}

\usepackage{graphicx}
\usepackage{multirow}
\usepackage{txfonts}
\usepackage{upgreek}    
\begin{document}

   \title{Asteroseismology of two $Kepler$ detached eclipsing binaries}

   \author{A. Liakos
           }
   \institute{Institute for Astronomy, Astrophysics, Space Applications and Remote Sensing, National Observatory of Athens,\\
            Metaxa \& Vas. Pavlou St., GR-15236, Penteli, Athens, Greece\\\\
              \email{alliakos@noa.gr}
             }

   \date{Received September XX, 2020; accepted March XX, 2020}


\abstract
{The present work contains light curve, spectroscopic and asteroseismic analyses for KIC~04851217 and KIC~10686876. These systems are detached eclipsing binaries hosting a pulsating component of $\delta$~Scuti type and have been observed with the unprecedented accuracy of the $Kepler$ space telescope. Using ground based spectroscopic observations, the spectral types of the primary components of the systems were estimated as A6V and A5V for KIC~04851217 and KIC~10686876, respectively, with an uncertainty of one subclass. The present spectral classification together with literature radial velocities curves were used to model the light curves of the systems and, therefore, to calculate the absolute parameters of their components with higher certainty. The photometric data were analysed using standard eclipsing binary modeling techniques, while their residuals were further analysed using Fourier transformation techniques in order to extract the pulsation frequencies of their host  $\delta$~Scuti stars. The oscillation modes of the independent frequencies were identified using theoretical models of $\delta$~Scuti stars. The distances of the systems were calculated using the relation between the luminosity and the pulsation period for  $\delta$~Scuti stars. The physical and the oscillation properties of the pulsating components of these systems are discussed and compared with others of the same type. Moreover, using all the currently known cases of $\delta$~Scuti stars in detached binaries, updated correlations between orbital and dominant pulsation periods and between $\log g$ and pulsation periods were derived. It was found that the proximity of the companion plays significant role to the evolution of the pulsational frequencies.}


   \maketitle
%

\section{Introduction}
\label{sec:intro}

The $\delta$~Scuti stars are fast and multiperiodic oscillating variables. They pulsate in radial and low-order non-radial pulsations due to $\kappa$-mechanism \citep[c.f.][]{AER10, BAL15}. They also pulsate in high-order non-radial modes which may be attributed to the turbulent pressure in the Hydrogen convective zone as suggested by \citet{ANT14} and \citet{GRA15}. They have masses between 1.5-2.5~$M_{\sun}$ \citep{AER10}, their spectral types range between A-F, and they extend from the main-sequence dwarfs up to giants lying mostly inside the classical instability strip. \citet{UYT11} used a sample of 750 A-F type stars, observed by $Kepler$ mission \citep{BOR10, KOC10}, and found that $63\%$ are either $\delta$~Scuti or $\gamma$~Doradus or hybrid stars. \citet{BOW18}, based on two large $Kepler$ ensembles of $\delta$~Scuti stars, explored the relations between their pulsational and stellar parameters, as well as their distribution within the classical instability strip. \citet{ZIA19} determined a period-luminosity relation for $\delta$~Scuti stars using $GAIA$ parallaxes \citep{GAI16}, assuming that their dominant oscillation mode is radial. \citet{MUR19}, using $Gaia$ DR2 parallaxes of 15,000 A-F stars and their respective $Kepler$ light curves, derived a sample of 2000 genuine $\delta$~Scuti stars, based on which they recalculated the boundaries of the classical instability strip on the Hertzsprung-Russell diagram. \citet{ANT19} published the first asteroseismic results for $\delta$~Scuti and $\gamma$~Doradus stars observed by the Transiting Exoplanet Survey Satellite mission \citep[TESS;][]{RIC09, RIC15}. \citet{JAY20} identified approximately 8400 $\delta$~Scuti stars among the data of the All-Sky Automated Survey for Supernovae \citep[ASAS-SN;][]{SHA14} and, by using distances from $Gaia$ DR2 parallaxes, derived pulsation period-luminosity relations for pulsators that oscillate in fundamental and overtone modes in specific wavelength bands.

The eclipsing binaries (hereafter EBs) are ultimate tools for the estimation of the physical parameters (e.g. masses, radii, luminosities) and the evolutionary status of their components, especially, when their light curves (hereafter LCs) and radial velocities curves (hereafter RVs) are combined in the analysis. Moreover, another powerful tool of the EBs is the `Eclipse Timing Variations' (ETV) method, which allows to detect orbital period modulating mechanisms (e.g. mass transfer, tertiary component etc).

The topic of $\delta$~Scuti stars in EBs can be considered as very interesting because it includes two totally different astrophysical subjects. Specifically, it combines the light variability due to geometric reasons (i.e. eclipses) with the intrinsic variability due to oscillations of the stellar interior and atmosphere. Therefore, in general, the pulsating stars in EBs are extremely important because their absolute parameters and their pulsating properties can be directly derived and compared with the respective theoretical models in order to check the validity of the stellar evolution theory. Moreover, the study of cases of close binaries (i.e. short period EBs) with strong interactions and mass transfer opens a new path to the exploration of the influence of the proximity effects due to binarity in the stellar interiors.

\begin{table*}
\centering
\caption{Known cases of binaries with a $\delta$~Scuti component observed by the $Kepler$ and $K2$ missions categorized according to their Roche geometry.}
\label{tab:Kepler EBs}
\scalebox{0.9}{
\begin{tabular}{l cc l cc l cc l cc }
\hline											
KIC No	&	ToV	&	Ref.	&	KIC No	&	ToV	&	Ref. 	&	KIC No	&	ToV	&	Ref.&KIC No	&	ToV	&	Ref.	\\
\hline																	
\multicolumn{3}{c}{ Detached}					&	9592855	&	SB2+EB	&	14	&	6220497	&	EB	&	24	&	5783368	&	EB	&	30	\\
\cline{1-3}																							
3858884	&	SB2+EB	&	1	&	9651065	&	O--C	&	11	&	6669809	&	EB	&	25	&	5872506	&	el	&	30	\\
4142768	&	SB2+EB	&	2	&	9851944	&	SB2+EB	&	15	&	8553788	&	SB1+EB	&	26	&	6381306	&	SB3	&	28, 29	\\
4150611	&	SB2+EB	&	3	&	10080943	&	SB2+el	&	16	&	8840638	&	EB	&	27	&	6541245	&	el	&	30	\\
4544587	&	SB2+EB	&	4	&	10661783	&	SB2+EB	&	17	&	10581918	&	SB2+EB	&	25	&	7756853	&	SB2	&	28, 29	\\
4851217	&	SB2+EB	&	pw	&	10686876	&	SB1+EB	&	18, pw	&	10619109	&	SB2+EB	&	25	&	8975515	&	SB2	&	28, 30	\\
6629588	&	EB	&	5	&	10736223	&	SB2+EB	&	19	&	11175495	&	EB	&	25	&	9775454	&	SB1	&	28, 31	\\
\cline{7-9}																							
8087799	&	EB	&	6	&	10989032	&	SB1+EB	&	6	&	\multicolumn{3}{c}{ Unknown}					&	10537907	&	SB1	&	28, 32	\\
\cline{7-9}																							
8113154	&	EB	&	7	&	10990452	&	O--C	&	11	&	4480321	&	SB3	&	28, 29	&	11572666	&	SB2	&	28, 33	\\
8197761	&	SB1+EB	&	8, 9	&	11401845	&	EB	&	20	&	4570326	&	el	&	30	&	11973705	&	SB2+el	&	34	\\
8262223	&	SB2+EB	&	10	&	11754974	&	SB1+O--C	&	21	&	4739791	&	EB	&	31	&	202843107\tablefootmark{a}	&	EB	&	35	\\
8264492	&	O--C	&	11	&	201534540\tablefootmark{a,}\tablefootmark{b}	&	SB1+EB	&	22	&	5197256	&	EB	&	32	&	245932119\tablefootmark{a}	&	EB	&	36	\\
\cline{4-6}																							
8569819	&	E	&	12	&	\multicolumn{3}{c}{Semi detached}					&	5219533	&	SB3	&	28, 29	&	246899376\tablefootmark{a,}\tablefootmark{c}	&	SB2+EB	&	37	\\
\cline{4-6}																							
9285587	&	SB1+EB	&	13	&	6048106	&	EB	&	23	&	5709664	&	SB2	&	33	&		&		&		\\
\hline						
\end{tabular}}
\tablefoot{SB1/2/3= Single (1), double (2), triple (3) line spectroscopic binary, EB=Eclipsing binary, el=ellipsoidal variable, O-C=Binarity confirmed through O-C analysis. \tablefoottext{a}{EPIC No}, \tablefoottext{b}{HD 099458}, \tablefoottext{c}{V1178~Tau}
\tablebib{(1) \citet{MAC14}, (2) \citet{GUO19}, (3) \citet{HEL17}, (4) \citet{HAM13}, (5) \citet{LIAN16}, (6) \citet{ZHA17}, (7) \citet{ZHA19}, (8) \citet{SOW17}, (9) \citet{GAU19}, (10) \citet{GUO17}, (11) \citet{MUS15}, (12) \citet{KUR15}, (13) \citet{FAI15}, (14) \citet{GUO17b}, (15) \citet{GUO16}, (16) \citet{SCH16}, (17) \citet{LEH13}, (18) \citet{LIAN17}, (19) \citet{CHE20}, (20) \citet{LEE17a}, (21) \citet{MUR13}, (22) \citet{SKA19}, (23) \citet{SAM18}, (24) \citet{Lee16a}, (25) \citet{LIA17}, (26) \citet{LIA18}, (27) \citet{YAN19}, (28) \citet{UYT11}, (29) \citet{LAM18}, (30) \citet{GAU14}, (31) \citet{LEE16b}, (32) \citet{TUR15}, (33) \citet{DER19}, (34) \citet{CAT11},  (35) \citet{OU19}, (36) \citet{LEE19b}, (37) \citet{SAN20}, (pw) present work.}}
\end{table*}

\begin{figure}
\includegraphics[width=\columnwidth]{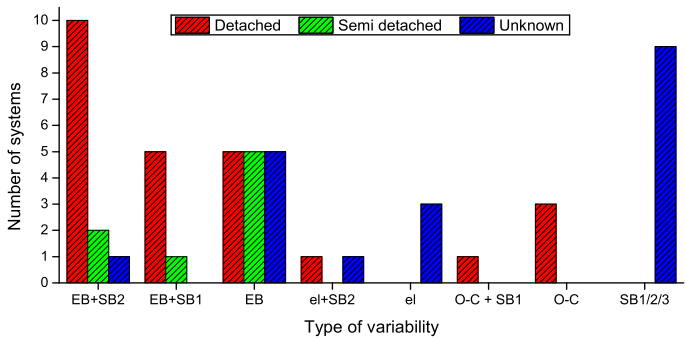}
\caption{Demographics of the published cases of binaries with a $\delta$~Scuti component observed by the $Kepler$ and $K2$ missions according to their Roche geometry and type of variability.}
\label{fig:KeplerStats}
\end{figure}

\citet{MKR02} introduced for first time the term `$oEA~stars$' (oscillating eclipsing binaries of Algol type) for characterizing the EBs with a $\delta$~Scuti mass accretor component of (B)A-F spectral type. The influence of mass accretion to the pulsational behaviour is one of the major questions of this particular topic of Asteroseismology and up to date there has been a lot discussion about it \citep[e.g.][]{TKA09, MKR18, BOW19}. \citet{SOY06a} announced the first connection between orbital ($P_{\rm orb}$) and dominant pulsation ($P_{\rm pul}$) periods for systems hosting a $\delta$~Scuti member. \citet{LIA12}, after a six year observational survey, published a catalogue including 74 cases and updated correlations between fundamental parameters. \cite{ZHA13} made the first theoretical attempt for the $P_{\rm pul}-P_{\rm orb}$ correlation. \citet{LIAN15, LIAN16} noticed for the first time a possible boundary between $P_{\rm pul}-P_{\rm orb}$ beyond that these two quantities can be considered uncorrelated ($P_{\rm orb}\sim13$~d). \citet{LIAN17} published the most complete catalogue for these systems to date (available online\footnote{\url{http://alexiosliakos.weebly.com/catalogue.html}}), which also includes updated correlations between fundamental parameters according to the geometrical status of the systems and the accuracy of their absolute parameters. A few months later, \citet{KAH17} almost doubled the threshold of $P_{\rm orb}$ by taking into account only the eclipsing systems. \citet{MUR18} published a review for binaries with pulsating components including also those with $\delta$~Scuti members. \citet{QIA18}, using the LAMOST spectra of more than 760 $\delta$~Scuti stars, found that 88 of them are possibly members of binary or multiple systems.

$Kepler$ \citep{BOR10, KOC10} and $K2$ \citep{HOW14} missions are fairly considered as a benchmark for asteroseismology, although the first one was designed for exoplanet transits. They have offered a great amount of high accuracy data (order of tenths of mmag) allowing the detection of extremely low amplitude frequencies \citep[order of a few $\upmu$mag;][]{MUR13, BOW18}. Moreover, the continuous recording of the targets eliminates the alias effects that are produced from the ground-based nightly-operated telescopes \citep{BRE00}. The time resolutions (short cadence $\sim1$~min; long cadence $\sim30$~min) of these space missions are extremely useful for the study of both EBs and short and long period pulsating stars. The easy access on the publicly available data from the $Kepler$ mission provided the means for astronomers to work at the same time in different topics. Especially for the EBs, a database hosting all the identified cases, namely `$Kepler$ Eclipsing Binary Catalog'\footnote{\url{http://keplerebs.villanova.edu/}} \citep[$KEBC$,][]{PRS11}, has been developed and provides the detrended data and preliminary information (e.g. ephemerides, magnitudes, temperatures) for a few thousands of EBs.

The catalogue of \citet{LIAN17} includes 17 cases of $Kepler$ EBs with a $\delta$~Scuti component. However, today, this sample has been tripled. In particular, up to date, there have been published in total eight oEA stars, 25 detached systems, and 19 systems with unclassified Roche geometry. All these systems are listed in Table~\ref{tab:Kepler EBs} along with their type of variability ($ToV$) and a corresponding reference, while their demographics are shown in Fig.~\ref{fig:KeplerStats}. For the majority of these systems (i.e. the eclipsing and the ellipsoidal cases), the absolute parameters of the pulsating component(s) have been calculated using either LCs only or both LCs and RVs.

The present paper is a continuation of the systematic study on EBs with a $\delta$~Scuti component \citep[see also][]{LIA09, LIA13, LIA14, LIAN16, LIA17, LIA18} and aims to enrich the current sample of this kind of systems. Preliminary results have been published for KIC~10686876 \citep{LIAN17}, regarding its dominant pulsational frequencies, and together with KIC~04851217 were selected for detailed analyses for first time. The selected sample of binaries consists the $\sim2.7\%$ of the 73 currently known detached binaries with a $\delta$~Scuti component \citep[][and subsequent individual publications as given in Table~\ref{tab:DEBsUPD}]{LIAN17} and the $\sim4.9\%$ of those with $P_{\rm orb}<13$~days. Moreover, the present sample increases the sample of such systems observed by $Kepler$ (i.e. with detailed asteroseismic modeling) by $\sim10.5\%$.

The results from the personal ground based spectroscopic observations, which allowed the estimation of the spectral types of the primary components of both studied systems, are given in Section~\ref{sec:sp}. Section~\ref{sec:LCmdl} includes the combined analysis of the $Kepler$ LCs and the spectroscopic data, which results in accurate modelings and estimations of the absolute parameters of the components of these systems. The further analyses of the LCs residuals, using Fourier transformation techniques, and the pulsational characteristics of the $\delta$~Scuti stars are presented in Section~\ref{sec:Fmdl}. The evolutionary status and the properties of the pulsating components of these systems are discussed and compared with other similar ones in Section~\ref{sec:Evol}. Discussion, summary and conclusions are presented in Section~\ref{sec:Dis}. 


\section{History of the selected systems}
\label{sec:history}

\subsection{KIC 04851217}
\label{sec:introKIC048}

This system (HD~225524) has a period of $\sim2.47$~d and a slightly eccentric orbit. Its first reference can be found in the Henry Draper extension catalogue \citep{CAN36}. Its first LC was obtained by \citet{HAR04} during the HATNET Variability Survey, while it was also observed by the ASAS survey \citep{PIG09}. \citet{ARM14} estimated the temperature of the system as 7022~K, while spectra obtained by LAMOST survey classified it between A5IV-A9V spectral types \citep{GRA16, FRA16, QIA18}. \citet{GIE15} calculated through an ETV analysis that the orbital period of the system increases with a rate $\sim1.71\times10^{-6}$~yr$^{-1}$. RVs for both components have been obtained by \citet[][$K_1$=115~km~s$^{-1}$, $K_2$=107~km~s$^{-1}$]{MAT17} and \citet[][$K_1$=131~km~s$^{-1}$ and $K_2$=115~km~s$^{-1}$]{HEL19}. In the latter work, the masses of the components were calculated based also on the $Kepler$ LCs.

\subsection{KIC 10686876}
\label{sec:introKIC106}
KIC~10686876 (TYC~3562-961-1) has an orbital period of $\sim2.62$~d. Its first LC was obtained by the Trans-atlantic exoplanet survey \citep[TrES;][]{DEV08}, while its temperature has been determined in the range 7944-8167~K \citep{SLA11, CHR12, HUB14}. \citet{GIE15}, \citet{ZAS15}, and \citet{BOR16} performed ETV analysis of the system and found a long term periodic term, which was attributed to the existence of a tertiary component. \citet{ZAS15} determined the period of the third body as $\sim6.7$~yr and its mass function as $f$($m_3$)=0.021~$M_{\sun}$. In the same study, a preliminary LC modeling was also performed assuming a mass ratio of 1, resulting in a third light contribution of $\sim5\%$. \citet{DAV16} listed it among the $Kepler$ systems exhibiting flare activity. \citet{LIAN17} included the system in their catalogue, mentioning also the dominant frequency of its $\delta$~Scuti component (21.02~d$^{-1}$). \citet{MAT17} performed spectroscopic observations and measured the RVs of the primary component ($K_1$=67~km~s$^{-1}$).


\section{Ground based Spectroscopy}
\label{sec:sp}

The spectroscopic observations aimed to derive the spectral types of the primary components of the selected systems. The spectra of the KIC targets were obtained with the 2.3~m Ritchey-Cretien `Aristarchos' telescope at Helmos Observatory in Greece in August and October 2016. The \emph{Aristarchos~Transient~Spectrometer}\footnote{\url{http://helmos.astro.noa.gr/ats.html}} (ATS) instrument \citep{BOU04} using the low resolution grating (600~lines~mm$^{-1}$) was employed for the observations. This set-up provided a resolution of $\sim3.2$~{\AA}~pixel$^{-1}$ and a spectral coverage between approximately 4000-7260~{\AA}. The spectral classification is based on a spectral line correlation technique between the spectra of the variables and standard stars, which were observed with the same set-up. In Table~\ref{tab:SpecResults} are listed: The name of the system (KIC), the dates of observations, the exposure time used (E.T.), and the orbital phase of the system when the spectrum was obtained ($\Phi_{\rm orb}$). All spectra were calibrated (bias, dark, flat-field corrections) using the \textsc{MaxIm DL} software. The data reduction (wavelength calibration, cosmic rays removal, spectra normalization, sky background removal) was made with the \textsc{RaVeRe} v.2.2c software \citep{NEL09}.

The correlation method was presented in detail in \citet{LIA17}, but is described here also in brief. The Balmer and the strong metallic lines between 4000-6800~{\AA} were used for the comparison between the spectra of the KIC systems and the standards. The differences of spectral line depths between each standard star and the variables derive sums of squared residuals in each case, with the least squares sums to indicate the best fit. This method works very well in cases of binaries with large luminosity difference between the components, where the primary star dominates the total spectrum. However, in order to check possible contribution of the secondaries to the observed spectra, which would lead to an underestimation of the temperatures of the primary stars, another method was applied. Specifically, using the standard stars spectra, all possible combinations (i.e. combined spectra; sum of spectra) between spectral types A, F, G, and K were calculated. Moreover, for each spectra combination, the individual spectra were given weights, that are related to the light contribution of the components. The starting value for the contribution of the primary component was 0.5 and the step was 0.05. For each spectra combination, ten sub-combinations with different light contributions of the components were calculated. Each combined spectrum was compared with those of the variables deriving sums of squared residuals. Similarly to the direct comparison method, the least squares sums lead to the best match.

\begin{figure}
\includegraphics[width=\columnwidth]{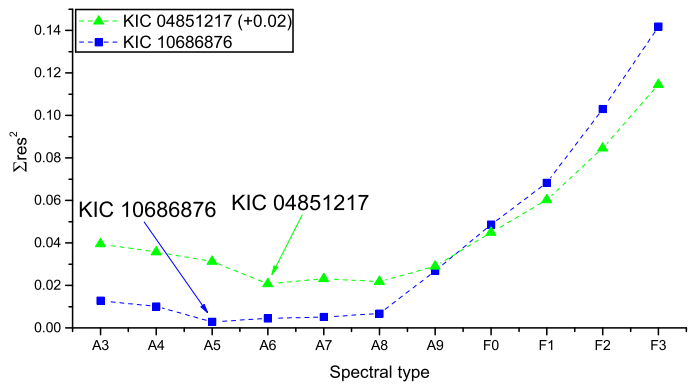}
\caption{Spectral type-search plots for KIC~04851217 and KIC~10686876. The points of KIC~04851217 are shifted vertically for better viewing. The arrows show the adopted spectral type for the primary component of each system. The comparison is shown only between A3-F3 spectral types due to scaling reasons.}
\label{fig:STs}
\end{figure}

\begin{figure}
\begin{tabular}{c}
\includegraphics[width=\columnwidth]{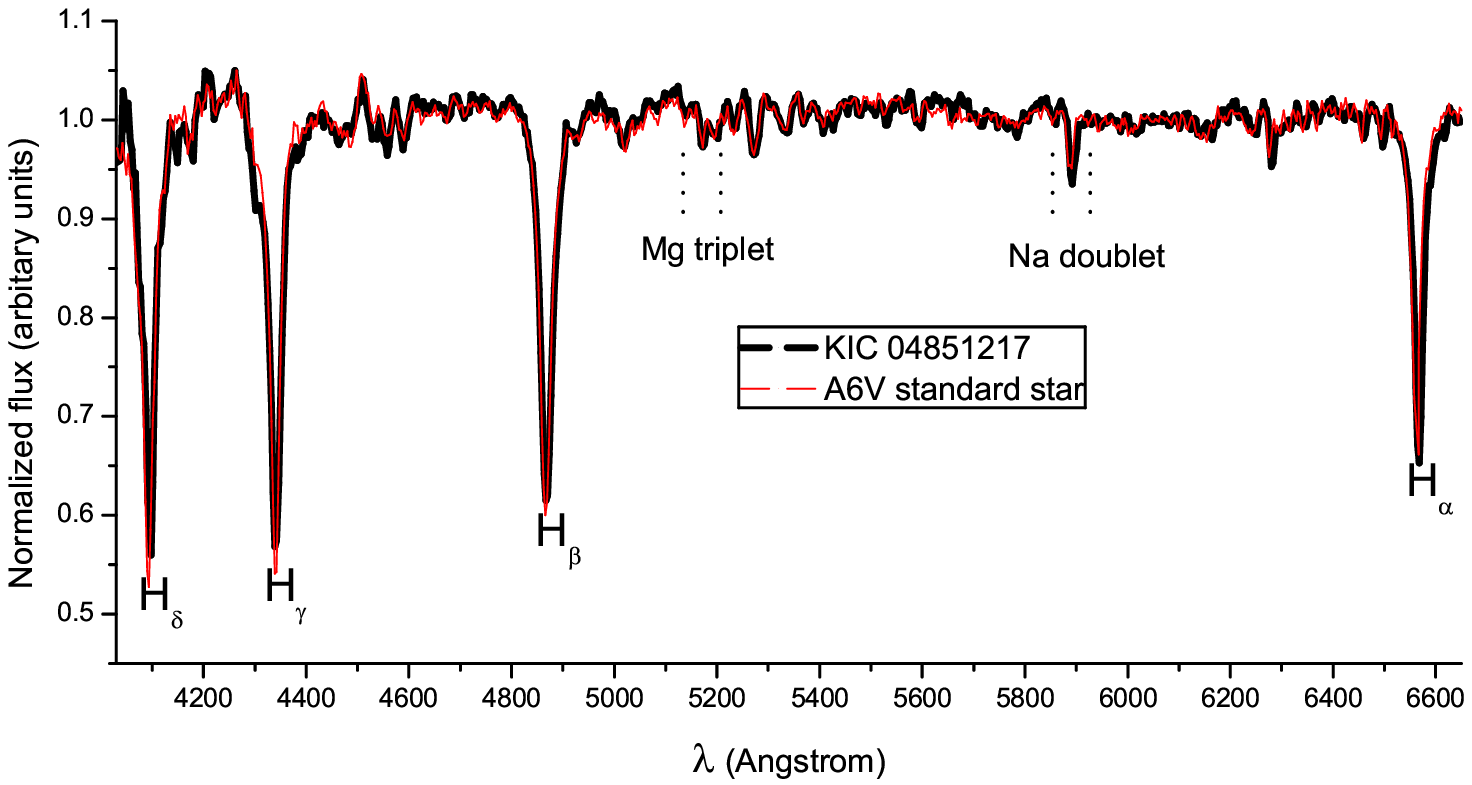}\\
\includegraphics[width=\columnwidth]{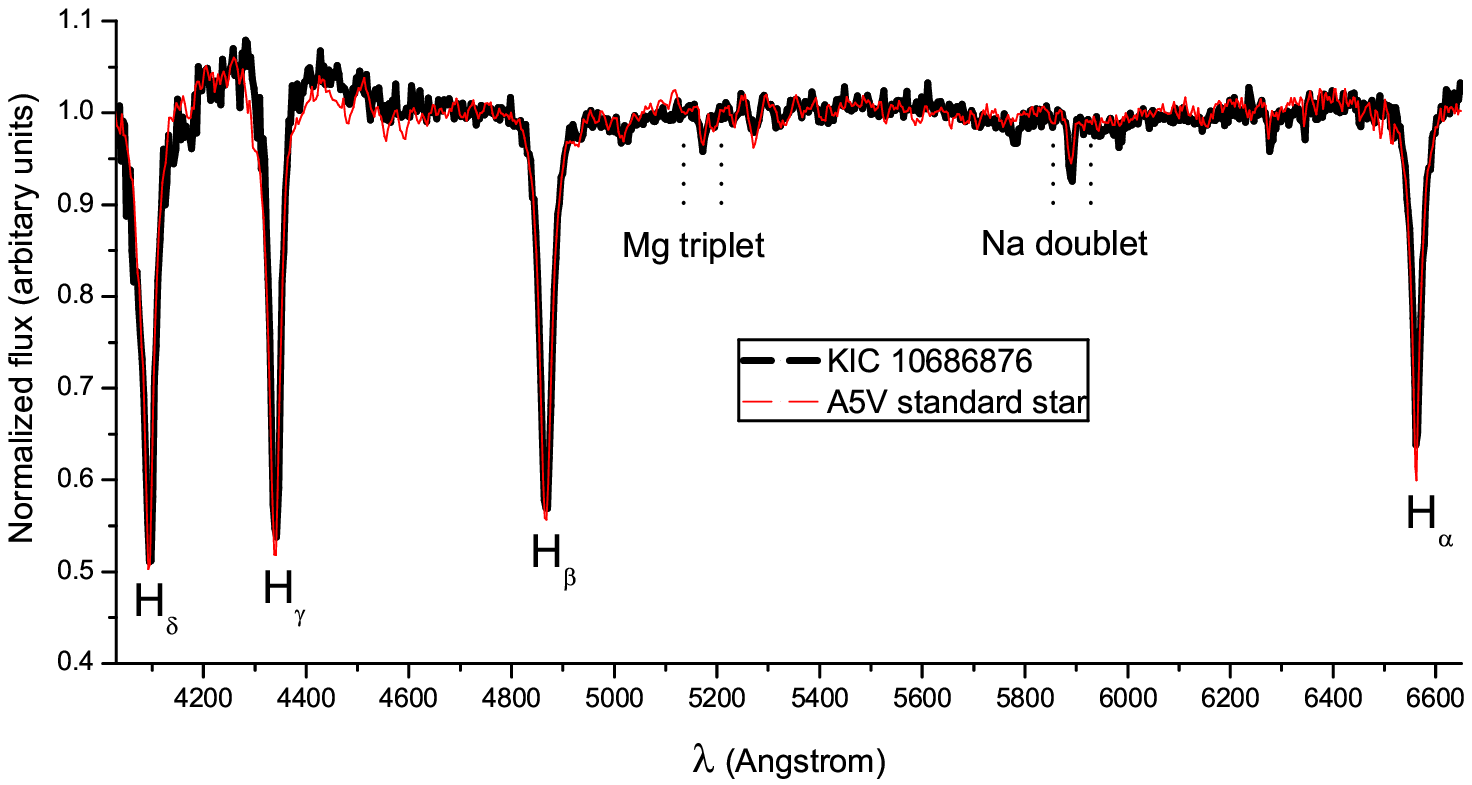}
\end{tabular}
\caption{Comparison spectra of the studied systems (black lines) and standard stars (red lines) with the closest spectral types. The Balmer and some strong metallic lines are also indicated.}
\label{fig:spectra}
\end{figure}

The light contribution of the primary of KIC~10686876 was found more than $95\%$, hence, its spectrum was compared directly with those of the standards. The components of KIC~04851217 were found to contribute almost equally to the observed spectrum, and, moreover, to be of the same spectral type. Therefore, practically, the combined spectrum was the same with that of a single star. The sum of squared residuals against the spectral type for these systems is plotted in Fig.~\ref{fig:STs}. The best fit for the spectrum of KIC~04851217 was found with that of an A6V standard star, and that of KIC~10686876 with an A5V standard star. The spectra of both systems along with those of best-fit standard stars are given in Fig.~\ref{fig:spectra}. Table~\ref{tab:SpecResults} includes also the spectral classification of the primary component of each system and the corresponding temperature values ($T_{\rm eff}$) according to the relations between effective temperatures and spectral types of \citet{COX00}. A formal error of one sub-class is adopted and is propagated also in the temperature determination. The present results come in very good agreement with those of previous studies within the error ranges (see Sections~\ref{sec:introKIC048}-\ref{sec:introKIC106}).

\begin{table}
\centering
\caption{Spectroscopic observations log and results for the primary components of the studied systems. The errors are given in parentheses alongside values and correspond to the last digit(s).}
\label{tab:SpecResults}
\begin{tabular}{l c cc cc}
\hline											
KIC No	&	Date	&	       E.T.	&	$\Phi_{\rm orb}$	&	Spectral 	&	$T_{\rm eff}$ 	\\
	&		    &	      (min)	&		               &	type	&	(K)	\\
\hline											
04851217	&	17/8/16	&	        10	&	0.67	&	A6V	&	8000(250)	\\
10686876	&	16/8/16	&	    10	    &	0.22	&	A5V	&	8200(250)	\\
\hline											
\end{tabular}
\end{table}

\section{Light curve modelling and absolute parameters calculation}
\label{sec:LCmdl}
\begin{figure*}[t]
\begin{tabular}{c}
\includegraphics[width=17.2cm]{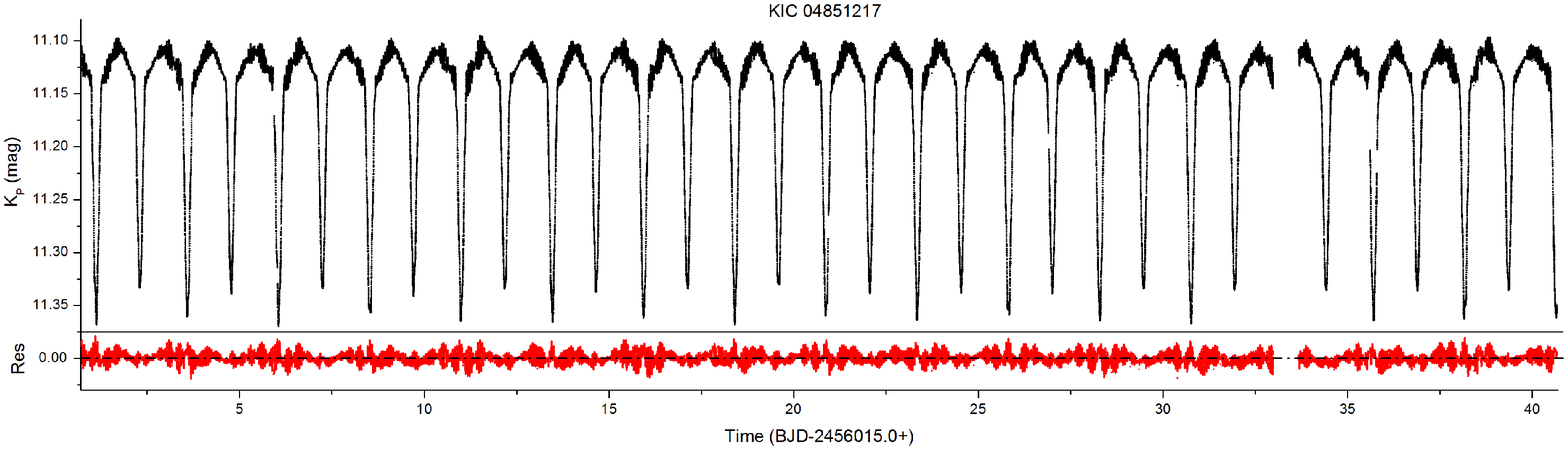}\\
\includegraphics[width=17.2cm]{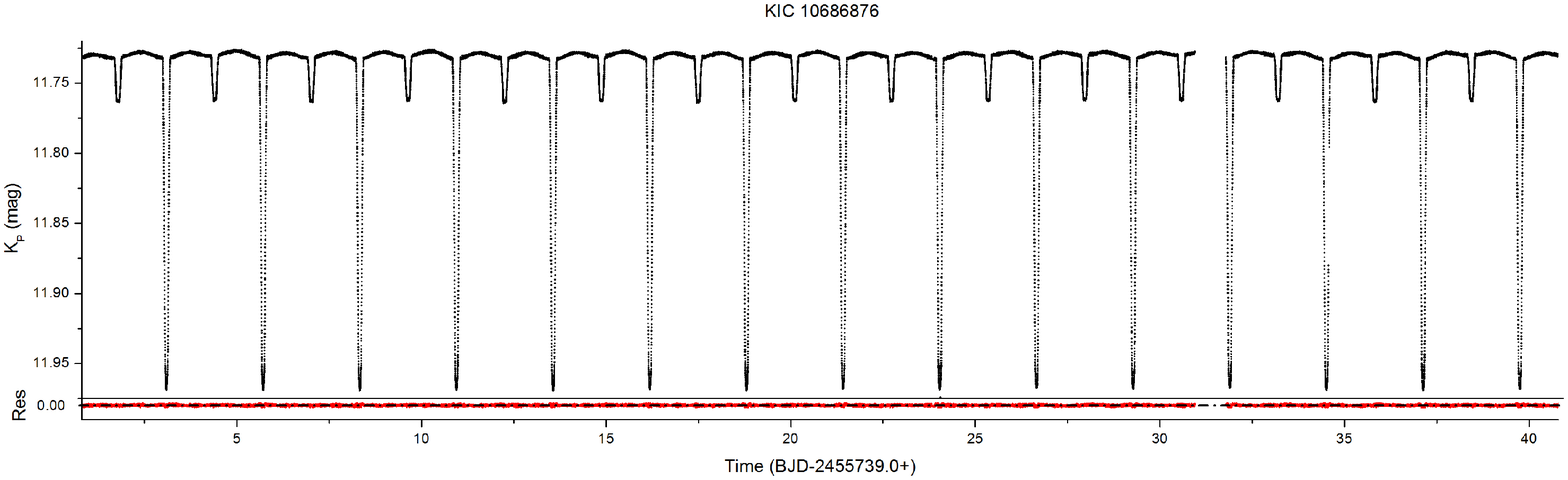}\\
\end{tabular}
\caption{Short cadence LCs (black points) for the studied systems and their residuals (red points) after the subtraction of the LC models. The plotted data concern only 40 days of observations, while the rest are not shown for scaling reasons.}
\label{fig:LCsandRes}
\end{figure*}

The selected systems were observed during many quarters of the $Kepler$ mission in long cadence mode. Additionally, both of them were also observed in short cadence mode during two or more quarters. Given that the primary objective of this work is the asteroseismic analysis of the pulsating stars of these systems (i.e. derivation of their oscillation frequencies and mode identification), only the short cadence data \citep[downloaded from the $KEBC$;][]{PRS11} were used for the subsequent analyses. The log of observations is given in Table~\ref{tab:Obslog}, which includes: The level of light contamination ($light~contam.$), the total points, the number of days that data cover, and the number of the fully covered LCs. For both systems, the total covering time of observations is more than 1.8 years, which can be considered extremely sufficient for the study of short-period pulsations (i.e. order of hours) and more than enough for LC modelling. The short cadence $Kepler$ data of approximately 40 days of observations for each system are plotted in Fig.~\ref{fig:LCsandRes}. The phases and the flux to magnitude conversions of both systems were calculated using the ephemerides ($T_{0}, P_{\rm orb}$) and the $K_{\rm p}$ magnitudes, respectively, as given in $KEBC$ \citep[][see also Table~\ref{tab:LCmdlAbs}]{PRS11}.

\begin{table}
\centering
\caption{Log of short cadence observations from the $Kepler$ mission for the studied systems.}
\label{tab:Obslog}
\scalebox{0.85}{
\begin{tabular}{l ccccc}
\hline													
KIC	&	Short cadence 	&	$light~contam.$	&	Points	&	days	&	LCs	\\
	&		quarter&	($\times10^{-3}$)	&	($\times10^5$)	&		&		\\
\hline											
04851217	&	2, 4, 9, 13, 15-17	&	3	&	6.21	&	1360	&	164	\\
10686876	&	3, 10	&	17	&	1.70	&	677	&	46	\\
\hline													
\end{tabular}}
\end{table}

The LCs analyses were made with the \textsc{PHOEBE} v.0.29d software \citep{PRS05} that is based on the 2003 version of the Wilson-Devinney code \citep{WIL71, WIL79, WIL90}. Temperatures of the primaries were assigned values derived from the present spectral classification (see Section~\ref{sec:sp} and Table~\ref{tab:SpecResults}) and were kept fixed during the analysis, while the temperatures of the secondaries $T_{2}$ were adjusted. Initially, for both systems, the albedos, $A_1$ and $A_2$, and gravity darkening coefficients, $g_1$ and $g_2$, were given values according to the spectral types of the components \citep{RUC69, ZEI24, LUC67}. The (linear) limb darkening coefficients, $x_1$ and $x_2$, were taken from the tables of \citet{HAM93}. The synchronicity parameters $F_1$ and $F_2$ were initially left free, but since they did not show any significant change during the iterations, they were given a value of 1 (tidal locking). The dimensionless potentials $\Omega_{1}$ and $\Omega_{2}$, the fractional luminosity of the primary component $L_{1}$, and the inclination of the system $i$ were set as adjustable parameters. The third light contribution $l_3$ was also set as adjustable parameter for the system which either has significant light contamination or there is information about possible tertiary component. For the present sample of systems, $l_3$ was taken into account only for KIC~10686876 (see Section~\ref{sec:KIC106mdldetails} for details), while for KIC~04851217 the light contamination is negligible ($<$0.3\%). At this point, it should to be noted that since the CCD sensors of $Kepler$ have effective wavelength response between approx. 410-910~nm with a peak at $\sim588$~nm, the $R$ filter (Bessell photometric system--range between 550-870~nm and with a transmittance peak at 597~nm) was selected as the best representative for the filter depended parameters (i.e. $x$ and $L$).

As described in the previous works \citep{LIA17, LIA18}, the analyses of EBs observed by $Kepler$ LCs need some special attention due to light variations between successive LCs. Therefore, in order to avoid a super simplified model including all data points folded into the $P_{\rm orb}$, the method of one model per LC (using normal points) or one model per a group of successive LCs folded into the $P_{\rm orb}$ was followed. The reasons for this are: a) the brightness changes from cycle to cycle due to magnetic activity and b) the asymmetries due to the pulsations (i.e. superimpositions or damping of the various frequency modes). It should be noticed that the derived LC residuals should exclude as much as possible the proximity and other stellar effects (e.g. magnetic activity) in order to perform subsequently an accurate frequency analysis (see next Section). The one model per LC or per a group of successive LCs technique provides also the means to derive more realistic errors for the adjusted parameters and minimize the widely known `unrealistic' error estimation of \textsc{PHOEBE}. The final LC model of a system includes the average values of the respective parameters from all the individual models, while their errors are the standard deviations of them. The LC analyses were applied in modes 2 (detached system), 4 (semi-detached system with the primary component filling its Roche lobe) and 5 (conventional semi-detached binary). Neither of these systems could converge in any semi-detached mode, therefore, both of them can be plausibly considered as detached EBs. The method described above is general and additional information for the analysis of each system is given in the following subsections.


\begin{table}
\centering
\caption{Modeling and absolute parameters for the studied systems. The errors are given in parentheses alongside values and correspond to the last digit(s).}
\label{tab:LCmdlAbs}
\begin{tabular}{l cc}
\hline					
System:	&	KIC 04851217	  &	KIC 10686876	\\
\hline					
	&  	\multicolumn{2}{c}{System parameters}			\\
\hline					
$K_{\rm p}^{(1)}$~(mag)	&  	11.11	&  	11.73	\\
$T_0^{(1)}$~(BJD)	&  	2454953.90(5)	&  	2454953.95(3)	\\
$P_{\rm orb}^{(1)}$~(d)	&  	2.470280(4)	&  	2.618427(4)	\\
$q$~($m_{2}$/$m_{1}$)	&  	1.14(4)	&  	0.44(1)	\\
$i$~($\degr$)	&  	76.7(1)	&  	86.4(1)	\\
$e$	&  	0.036(1)	&  	-	\\
$\omega$~($\degr$)	&  	2.64(1)	&  	-	\\
\hline					
	&  	\multicolumn{2}{c}{Components parameters}			\\
\hline					
$T_{1}$~(K)	&  	8000(250)$^{(2)}$	&  	8200(250)$^{(2)}$	\\
$T_{2}$~(K)	&  	7890(98)	&  	4740(115)	\\
$\Omega_1$	&  	5.94(1)	&  	6.14(1)	\\
$\Omega_2$	&  	6.26(1)	&  	7.00(8)	\\
$K_1$~(km~s$^{-1}$)	&  	131(2.6)$^{(3)}$	&  	67(2)$^{(4)}$	\\
$K_2$~(km~s$^{-1}$)	&  	114.6(2.7)$^{(3)}$	&  	-	\\
$A_1$	&  	1\tablefootmark{a}	&  	1\tablefootmark{a}	\\
$A_2$	&  	1\tablefootmark{a}	&  	1\tablefootmark{a}	\\
$g_1$	&  	1\tablefootmark{a}	&  	1\tablefootmark{a}	\\
$g_2$	&  	1\tablefootmark{a}	&  	0.32\tablefootmark{a}	\\
$x_1$	&  	0.439	&  	0.377	\\
$x_2$	&  	0.444	&  	0.500	\\
$L_1$/$L_{\rm T}$	&  	0.585(2)	&  	0.948(1)	\\
$L_2$/$L_{\rm T}$	&  	0.415(1)	&  	0.023(2)	\\
$L_3$/$L_{\rm T}$	&  	-	&  	0.029(1)	\\
$r_{1,~\rm pole}$	&  	0.209(1)	&  	0.175(1)	\\
$r_{2,~\rm pole}$	&  	0.217(1)	&  	0.077(1)	\\
$r_{1,~\rm point}$	&  	0.216(1)	&  	0.177(1)	\\
$r_{2,~\rm point}$	&  	0.223(1)	&  	0.077(1)	\\
$r_{1,~\rm side}$	&  	0.212(1)	&  	0.176(1)	\\
$r_{2,~\rm side}$	&  	0.219(1)	&  	0.077(1)	\\
$r_{1,~\rm back}$	&  	0.215(1)	&  	0.177(1)	\\
$r_{2,~\rm back}$	&  	0.222(1)	&  	0.077(1)	\\
\hline					
	&  	\multicolumn{2}{c}{Absolute parameters}			\\
\hline					
$M_1~$($M_{\sun}$)	&  	1.92(10)	&  	2.0(2)\tablefootmark{a}	\\
$M_2~$($M_{\sun}$)	&  	2.19(18)	&  	0.88(9)	\\
$R_1~$($R_{\sun}$)	&  	2.61(5)	&  	2.00(7)	\\
$R_2~$($R_{\sun}$)	&  	2.68(5)	&  	0.88(9)	\\
$L_1~$($L_{\sun}$)	&  	25(3)	&  	16(2)	\\
$L_2~$($L_{\sun}$)	&  	25(3)	&  	0.3(1)	\\
$\log g_1$~(cm~s$^{-2}$)	&  	3.89(3)	&  	4.14(5)	\\
$\log g_2$~(cm~s$^{-2}$)	&  	3.92(4)	&  	4.50(10)	\\
$a_1$~($R_{\sun}$)	&  	6.6(1)	&  	3.5(1)	\\
$a_2$~($R_{\sun}$)	&  	5.7(1)	&  	7.9(3)	\\
$M_{\rm bol,~1}$~(mag)	&  	1.26(6)	&  	1.73(8)	\\
$M_{\rm bol,~2}$~(mag)	&  	1.26(4)	&  	5.9(1)	\\
\hline																														
\end{tabular}
\tablefoot{$L_{\rm T}=L_1+L_2+L_3$.
\tablefoottext{a}{assumed}
\tablebib{(1) $KEBC$-\citet{PRS11}, (2)~Section~\ref{sec:sp}, (3)~\citet{HEL19}, (4)~\citet{MAT17}.}
}

\end{table}

\begin{figure}
\begin{tabular}{c}
\includegraphics[width=\columnwidth]{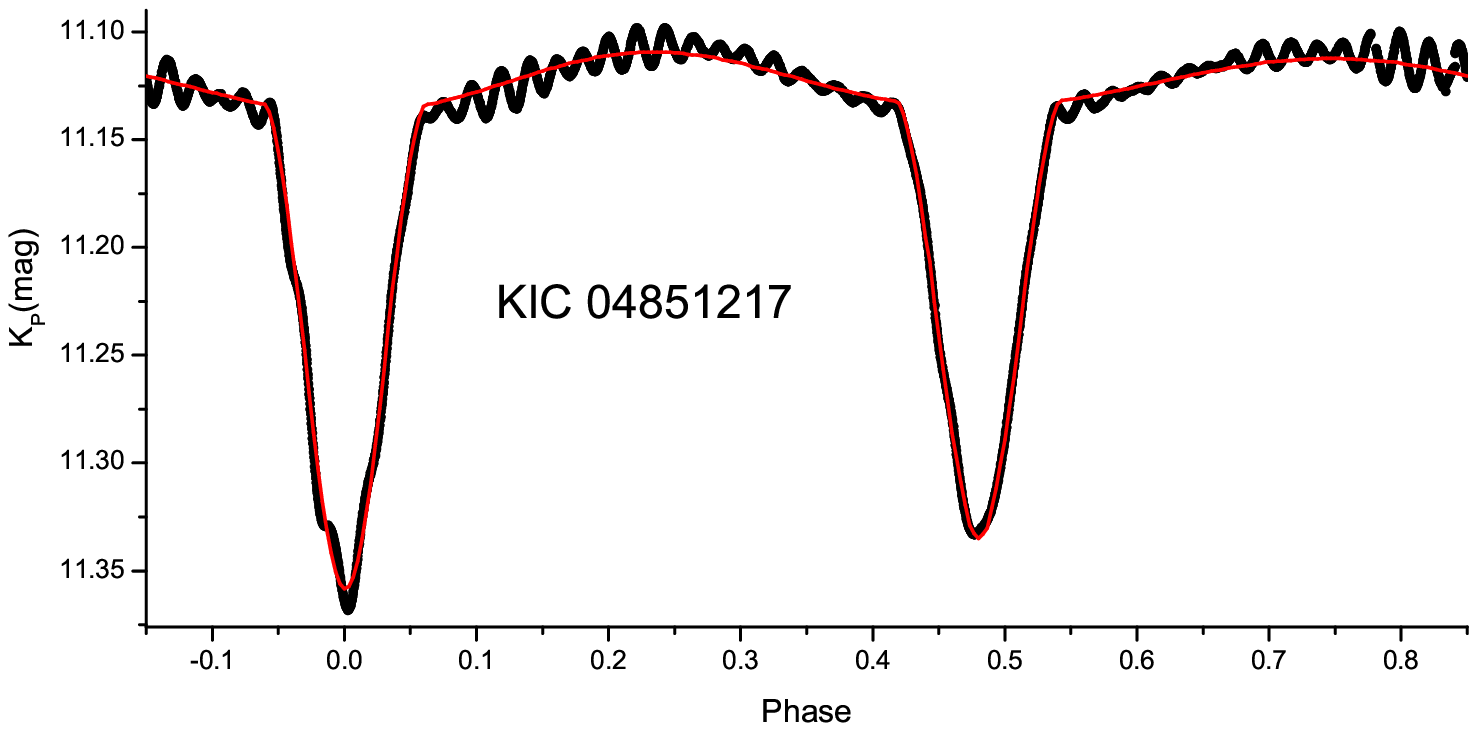}\\
\includegraphics[width=6cm]{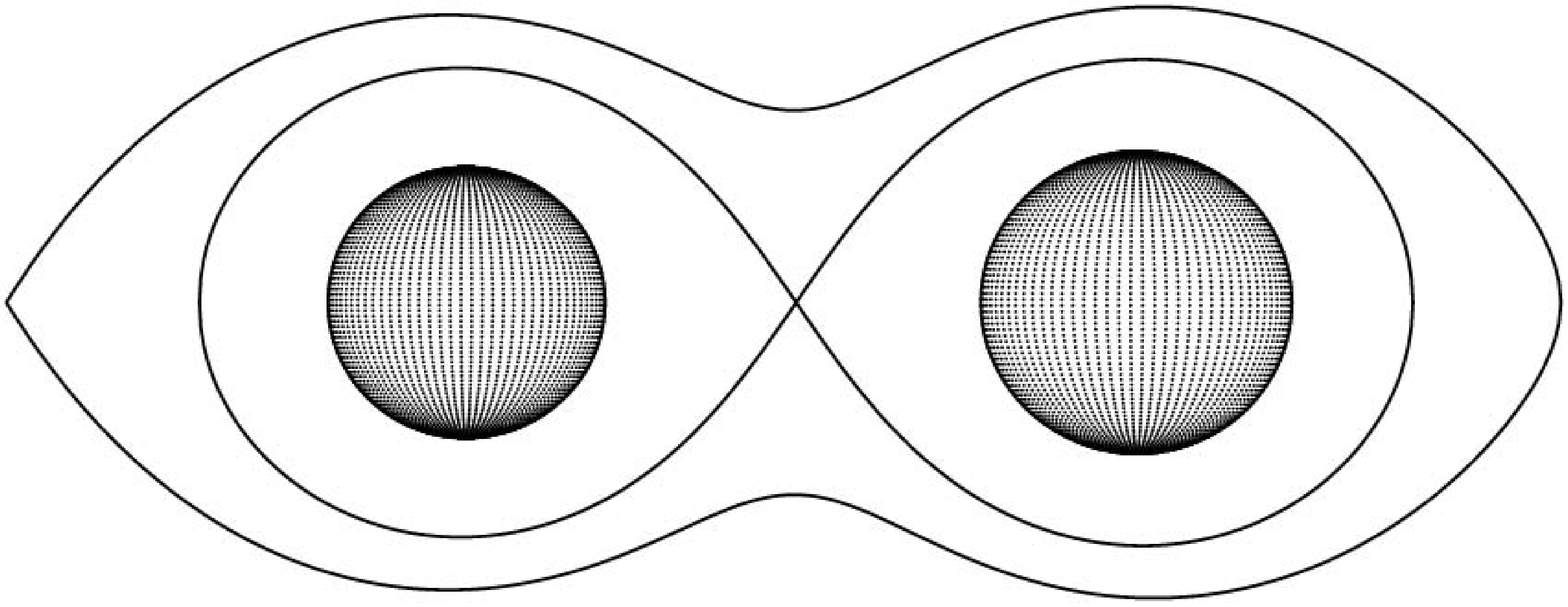}\\
\includegraphics[width=\columnwidth]{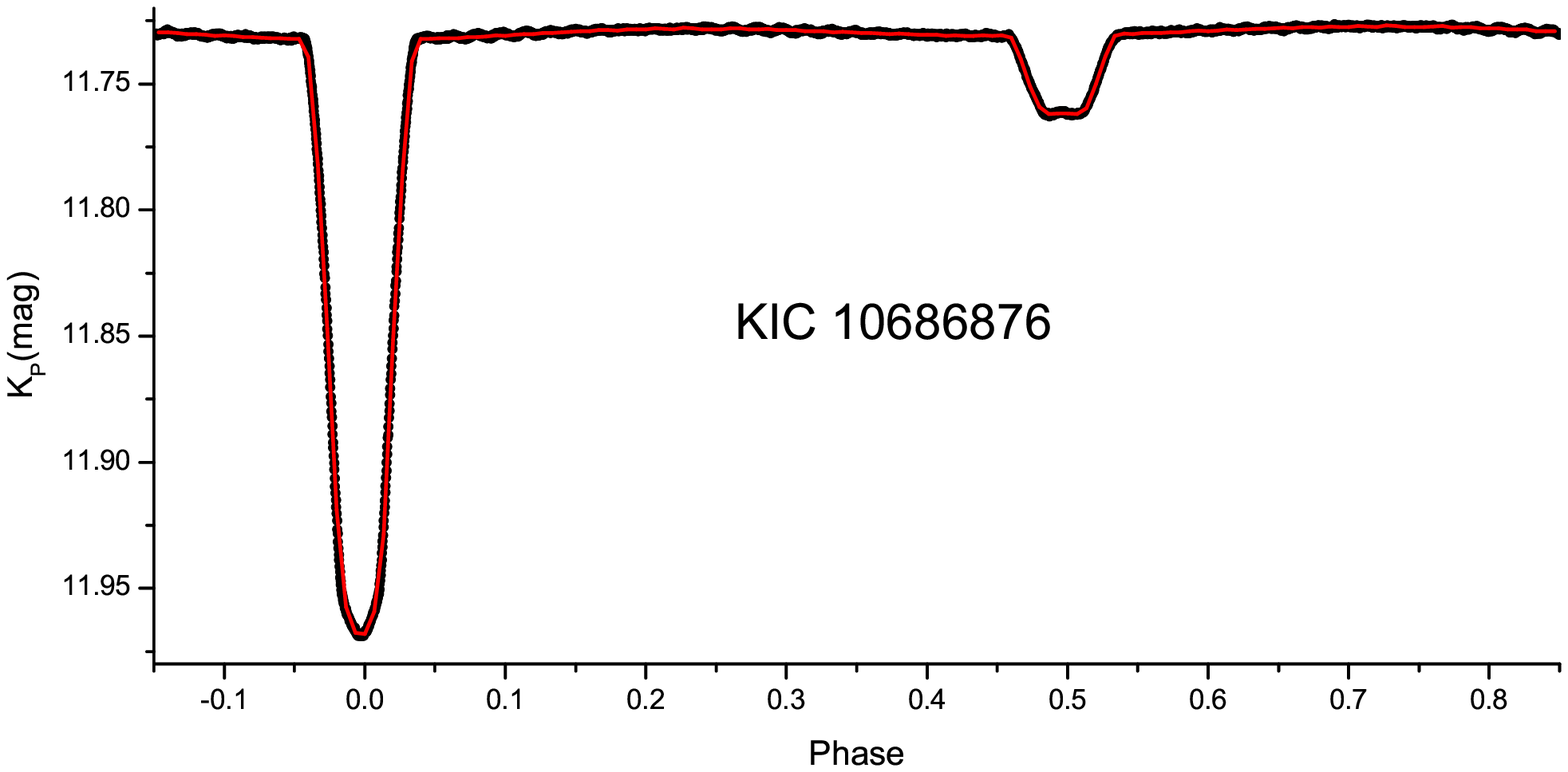}\\
\includegraphics[width=6cm]{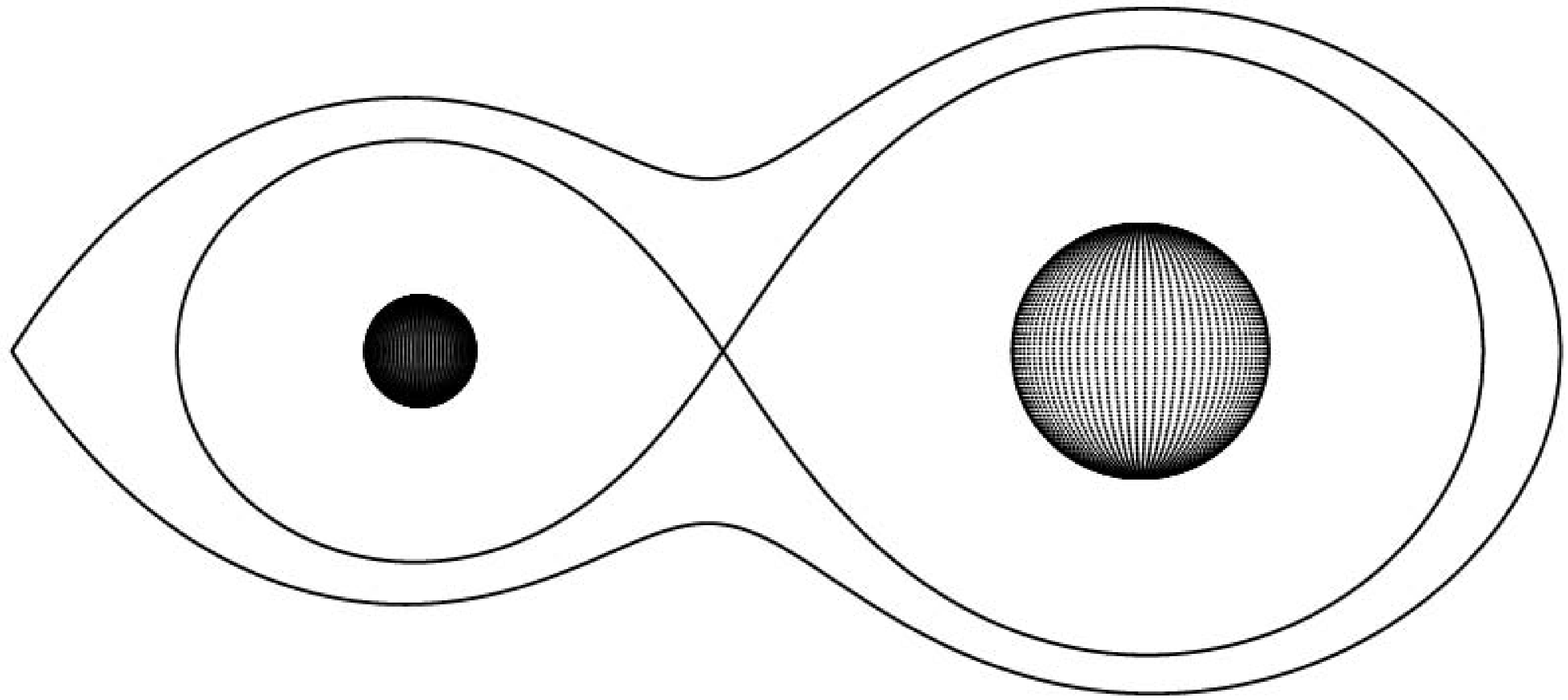}\\
\end{tabular}
\caption{Theoretical (solid lines) over observed (points) LCs (one cycle) and two-dimensional representations of the Roche geometry at orbital phase 0.75 for the studied systems.}
\label{fig:LCm3D}
\end{figure}

\subsection{KIC 04851217}
\label{sec:KIC048mdldetails}
For this system there are two spectroscopic studies regarding the RVs of its components in the literature. As mentioned in Section~\ref{sec:introKIC048}, \citet{MAT17} calculated the amplitudes of the RVs as $K_1$=$115$~km~s$^{-1}$ and $K_2$=107~km~s$^{-1}$, for the primary and the secondary components, respectively, while \citet{HEL19} found $K_1$=131~km~s$^{-1}$ and $K_2$=115~km~s$^{-1}$. In the latter study, the authors, additionally to their personal observations, they also used data from the Trans-atlantic Exoplanet Survey \citep[TrES;][]{ALO04} and Apache Point Observatory Galactic Evolution Experiment \citep[APOGEE;][]{MAJ17}. Therefore, that work is considered as more complete and reliable, given the larger data set used, hence, their results were adopted for the present study. According to these $K_1$ and $K_2$ values, a mass ratio ($q$) of 1.14$\pm$0.04 was calculated. This value was initially assigned in \textsc{PHOEBE}, but during the iterations it was left free to adjust inside its error range. Moreover, due to the displacement of the secondary minimum from the phase 0.5, the eccentricity $e$ and the argument of periastron $\omega$ were also left free to adjust. Given that no brightness changes due to magnetic activity were detected, one model per approximately six successive LCs folded into the $P_{\rm orb}$ was derived, resulting in total in 27 individual LC models.

\subsection{KIC 10686876}
\label{sec:KIC106mdldetails}
The analysis of this system was revealed as the most complicated in comparison with the previous one. First of all, using the findings of \citet{MAT17}, who measured the RV1 and calculated $K_1$=67$\pm$2~km~s$^{-1}$, and according to the present results about the spectral type of the primary component (A5V), the analysis was constrained in the range $0.4<q<0.5$ \citep[c.f.][]{WAN19}. In this regime, the expected mass of the primary ranges between 1.5-2.5~$M_{\sun}$. However, following exactly the same method likely for the previous EB regarding the fixed parameters and by adopting spots to fit the asymmetries, no sufficient solution was found in this $q$ range. Therefore, the iterations continued beyond this range and the best solution was found for $q\sim0.28$. The LC solution for this $q$ value derives $M_1$=6~$M_{\sun}$, $M_2$=1.72~$M_{\sun}$ and $T_2\sim4800$~K. Although, that was the best numerical solution based on the sum of squared residuals, however, there was not any physical meaning regarding the parameters of the components. Therefore, the modelling started again in the $q$ range 0.4-0.5, with the difference that $A_2$ was set to 1, which can be interpreted as reflection from the much hotter primary component. KIC~10686876 presents significant brightness changes due to magnetic activity \citep[listed also as system exhibiting flare activity by][]{DAV16}, hence, hot and cool spots on the surface of the secondary component were adopted. In addition, $l_3$ was also adjusted because the system a) has a light contamination of 1.7$\%$ (Table~\ref{tab:Obslog}) and b) possibly hosts a tertiary component \citep{GIE15, ZAS15, BOR16}.

The LCs modeling results for both systems are given in the upper and middle parts of Table~\ref{tab:LCmdlAbs}. Examples of LC modeling and the Roche geometry of each system are illustrated in Fig.~\ref{fig:LCm3D}.The LCs residuals after the subtraction of the individual models are shown below the observed LCs in Fig.~\ref{fig:LCsandRes}. In addition, the parameters of the spots for each LC (cycle) of KIC~10686876 (Table~\ref{tab:spots}) are given in appendix~\ref{sec:App2}. The variations of the parameters of the spots over time and various 2D representations showing the positions of the spots on the surface of the secondary component for two different dates of observations are plotted in Fig.~\ref{fig:spotKIC106}.

Using the RVs curves for these systems, the absolute parameters of their components can be fairly formed. However, given that for KIC~10686876 only the RV1 has been measured, the mass of its primary was inferred from its spectral type according to the spectral type-mass correlations of \citet{COX00} for main-sequence stars. A fair mass error value of 10$\%$ was also adopted. The secondary's mass follows from the determined mass ratio. The semi-major axes $a$, which fix the absolute mean radii, can then be derived from Kepler's third law. The luminosities ($L$), the gravity acceleration ($\log g$), and the bolometric magnitudes values ($M_{\rm bol}$) were calculated using the standard definitions. The absolute parameters were calculated with the software \textsc{AbsParEB} \citep{LIA15} and they are given in the lower part of Table~\ref{tab:LCmdlAbs}.


\section{Pulsation frequencies analysis}
\label{sec:Fmdl}

The results from spectroscopy and LCs modelling (Sections~\ref{sec:sp} and \ref{sec:LCmdl}) indicate that for KIC~10686876 only the primary component fits well to the profiles of the  $\delta$~Scuti type stars (i.e. mass and temperature). On the other hand, the components of KIC~04851217 show similar physical properties, therefore, additional search was needed in order to reveal the pulsating component (see Section~\ref{sec:KIC048freqdetails}). The pulsation frequencies analysis was performed with the software \textsc{PERIOD04} v.1.2 \citep{LEN05} that is based on classical Fourier analysis. The typical frequencies range of $\delta$~Scuti stars is 4-80~d$^{-1}$ \citep{BRE00, BOW18}, hence, the search should had been made for this range. However, an extensive range between 0-80~d$^{-1}$ was selected for the search, since  $\delta$~Scuti stars in binary systems may also present $g$-mode pulsations that are connected to their $P_{\rm orb}$ or present hybrid behaviour of $\gamma$~Doradus- $\delta$~Scuti type. For these analyses, the LCs residuals of each system (i.e. after the subtraction of the LCs models) were used (Fig.~\ref{fig:LCsandRes}). A 4$\upsigma$ limit \citep[S/N$=4$; recommended by][]{LEN05} regarding the reliability of the detected frequencies was adopted. 
Hence, after the first frequency computation the residuals were subsequently pre-whitened for the next one until the detected frequency had S/N$\sim$4. The Nyquist frequency and the frequency resolution according to the Rayleigh-Criterion (i.e. 1/$T$, where $T$ is the observations time range in days) are 733~d$^{-1}$ and 0.0007~d$^{-1}$ for KIC~04851217, and 732~d$^{-1}$ and 0.001~d$^{-1}$ for KIC~10686876, respectively.

After the frequencies search, the relation of \citet{BRE00} was used for the calculation of the pulsation constants ($Q$) of the independent frequencies:\\

\begin{equation}
\log Q = -\log f + 0.5 \log g + 0.1M_{\rm bol} + \log T_{\rm eff}- 6.456,\\
\label{eq:Q}
\end{equation}
where $f$ is the frequency of the pulsation mode, while $\log g$, $M_{\rm bol}$, and $T_{\rm eff}$ denote the standard quantities (see Section~\ref{sec:LCmdl}). Subsequently, using the pulsation constant - density relation:

\begin{equation}
Q = f_{\rm dom}^{-1}\sqrt{\rho_{\rm pul}/\rho_{\sun}},\\
\label{eq:rho}
\end{equation}
where $f_{\rm dom}$ is the frequency of the dominant pulsation mode (i.e. that with the largest amplitude), the density of the pulsator $\rho_{\rm pul}$ in solar units was also calculated. The results for the aforementioned quantities for each pulsator are given in Table~\ref{tab:IndF}. It should be noticed that the $Q$ values for all modes of both pulsators, except one for KIC~10686876, lie within the typical range of $\delta$~Scuti stars \citep[i.e. $0.015<Q<0.035$;][]{BRE00}. Moreover, it should to be noted that the majority of  $\delta$~Scuti stars exhibit intrinsic amplitude modulation \citep{BOW16, BAR20}, hence, $f_{\rm dom}$ may differ from time to time. Therefore, the present calculated densities might be not that accurate. However, they can be considered as rather representatives, since the respective calculations for the other pulsation modes for both systems yield very similar values.

The mode identification (i.e. $l$-degrees and the type of each oscillation) was made using the theoretical MAD models for  $\delta$~Scuti stars \citep{MON07} in the \textsc{FAMIAS} software v.1.01 \citep{ZIM08}. These models have a grid step 0.05-0.1~$M_{\sun}$ in the mass range 1.6-2.2~$M_{\sun}$, which is sufficient for the studied systems. The models ($l$-degrees) of the detected independent frequencies that corresponded to the parameters ($\log g$, $M$, and $T_{\rm eff}$) of the pulsators were adopted as the most probable oscillation modes. Moreover, $P_{\rm pul}$/$P_{\rm orb}$ ratios for all frequencies were also calculated and found below 0.07, except for $f_{7}$ of KIC~10686876, indicating that they are probably of $p$-mode type \citep{ZHA13}.

\begin{figure*}
\centering
\begin{tabular}{cc}
\includegraphics[width=8.5cm]{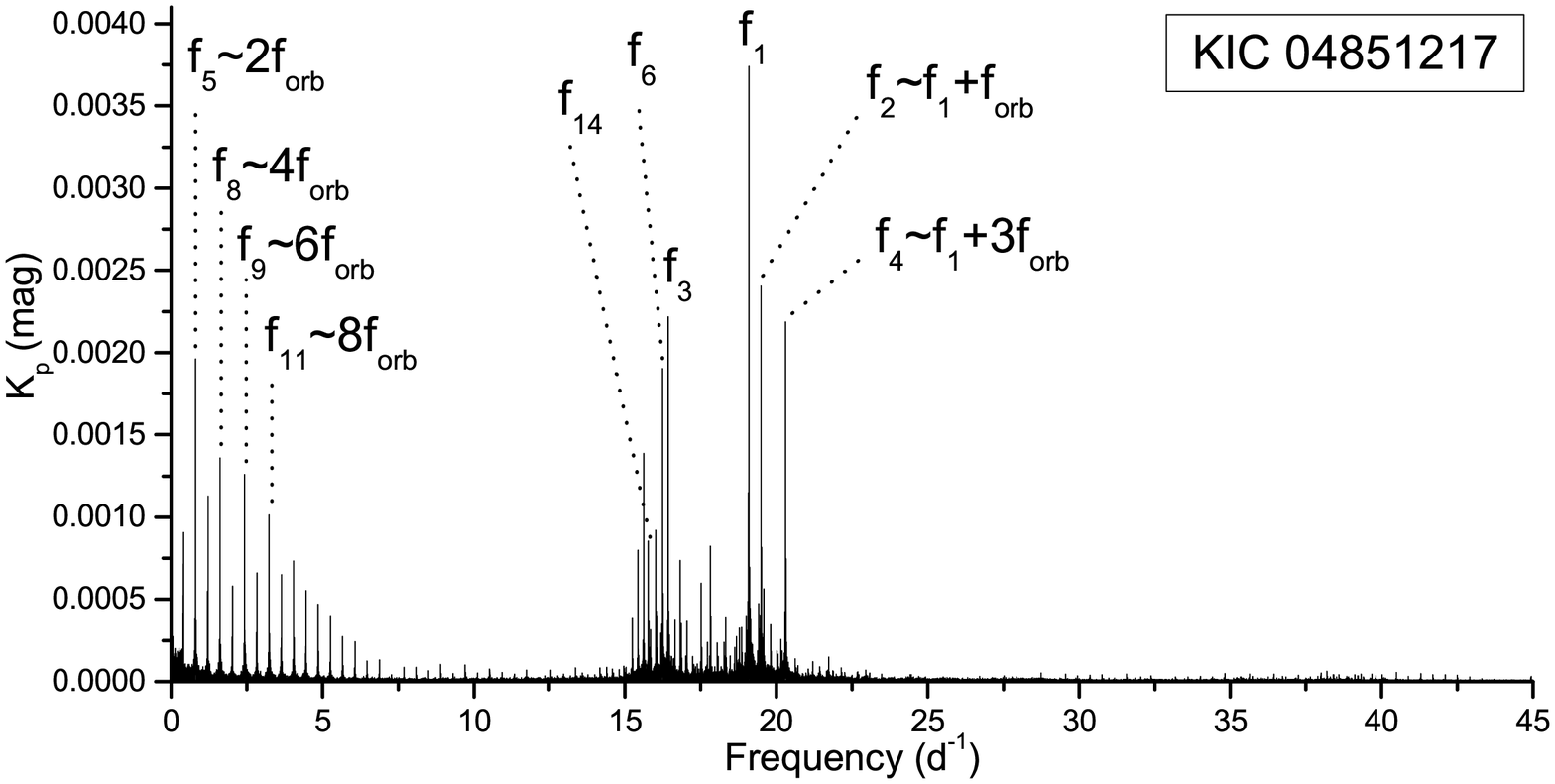}&\includegraphics[width=8.5cm]{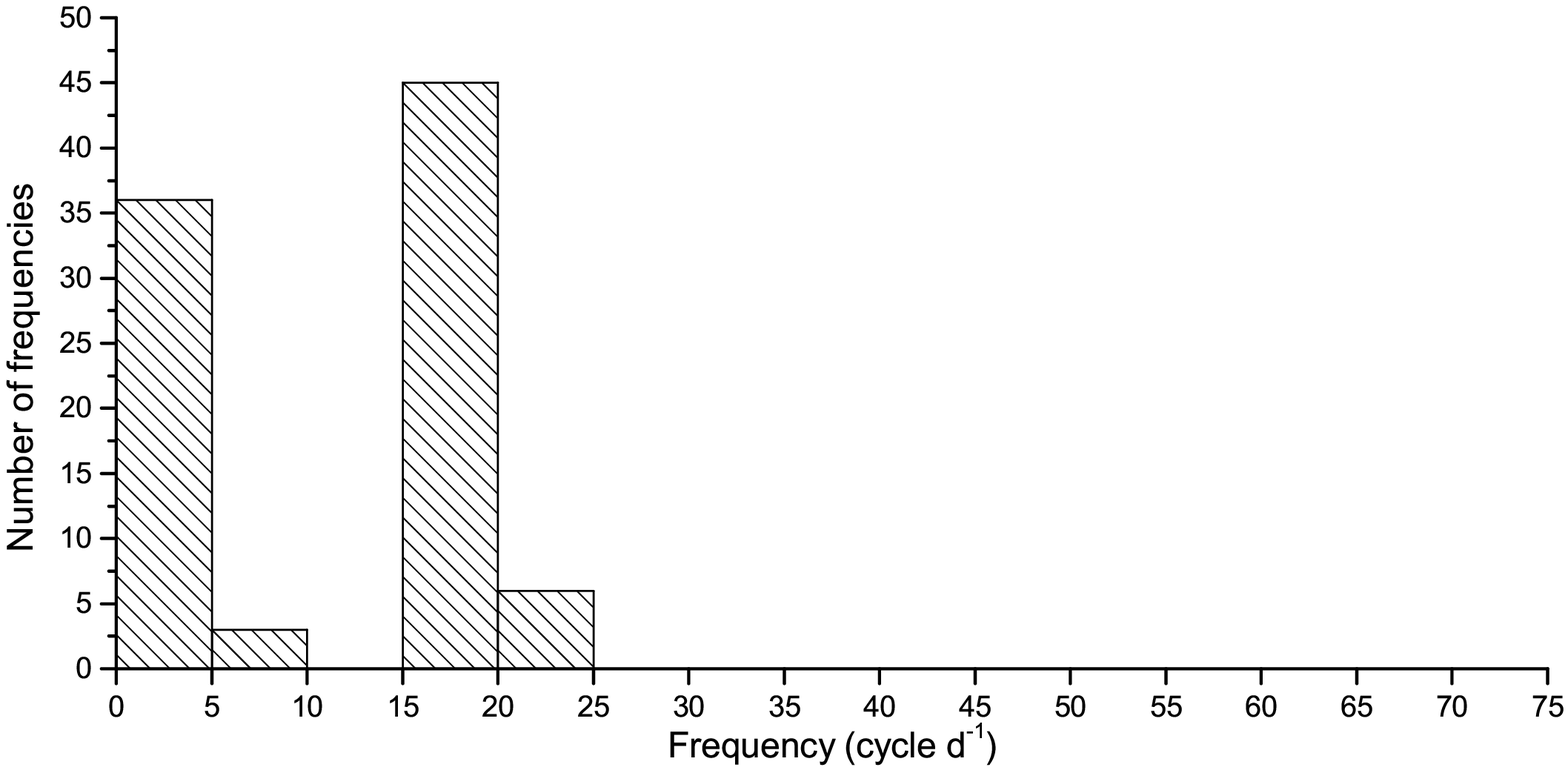}\\
\includegraphics[width=8.5cm]{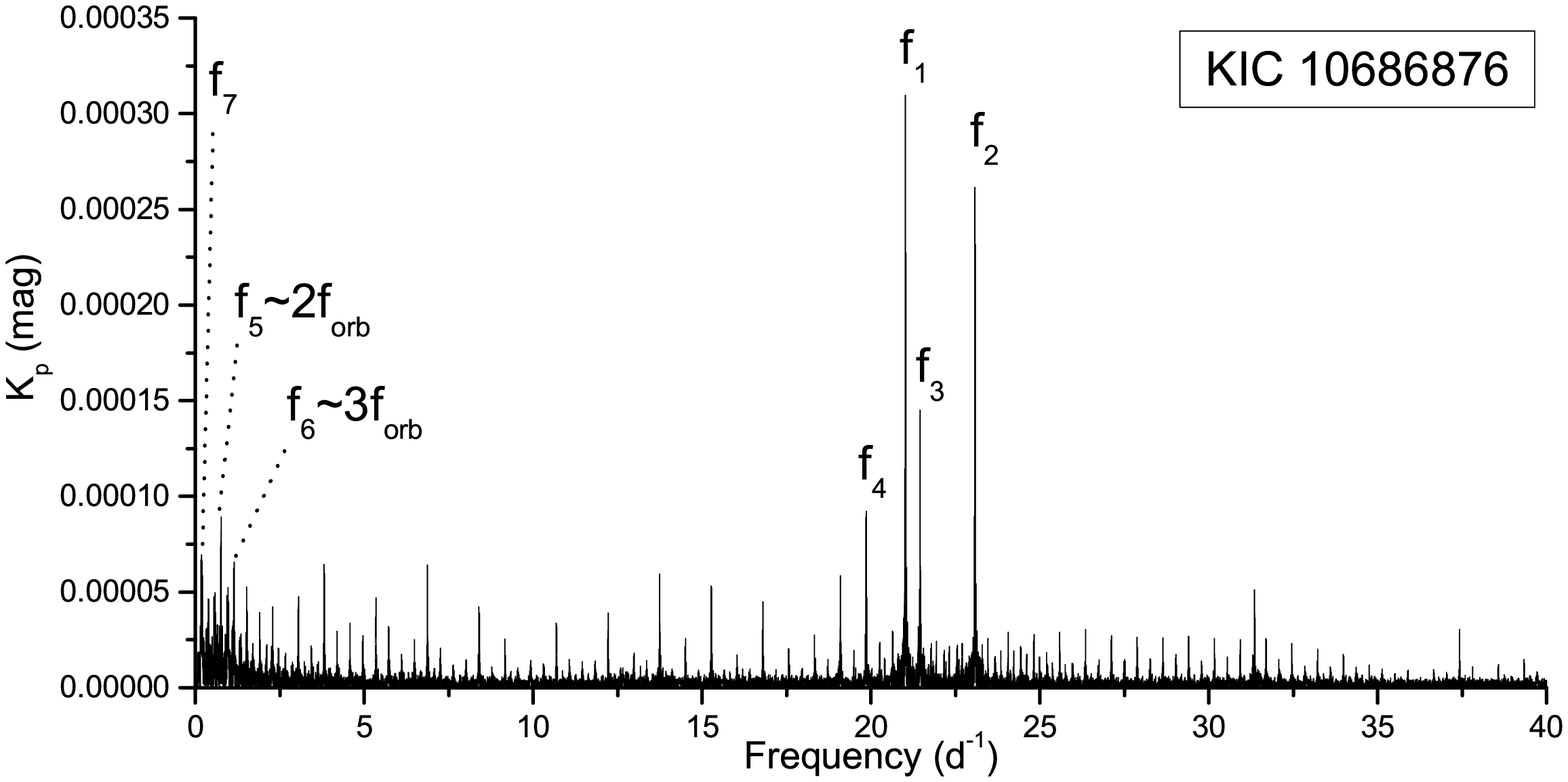}&\includegraphics[width=8.5cm]{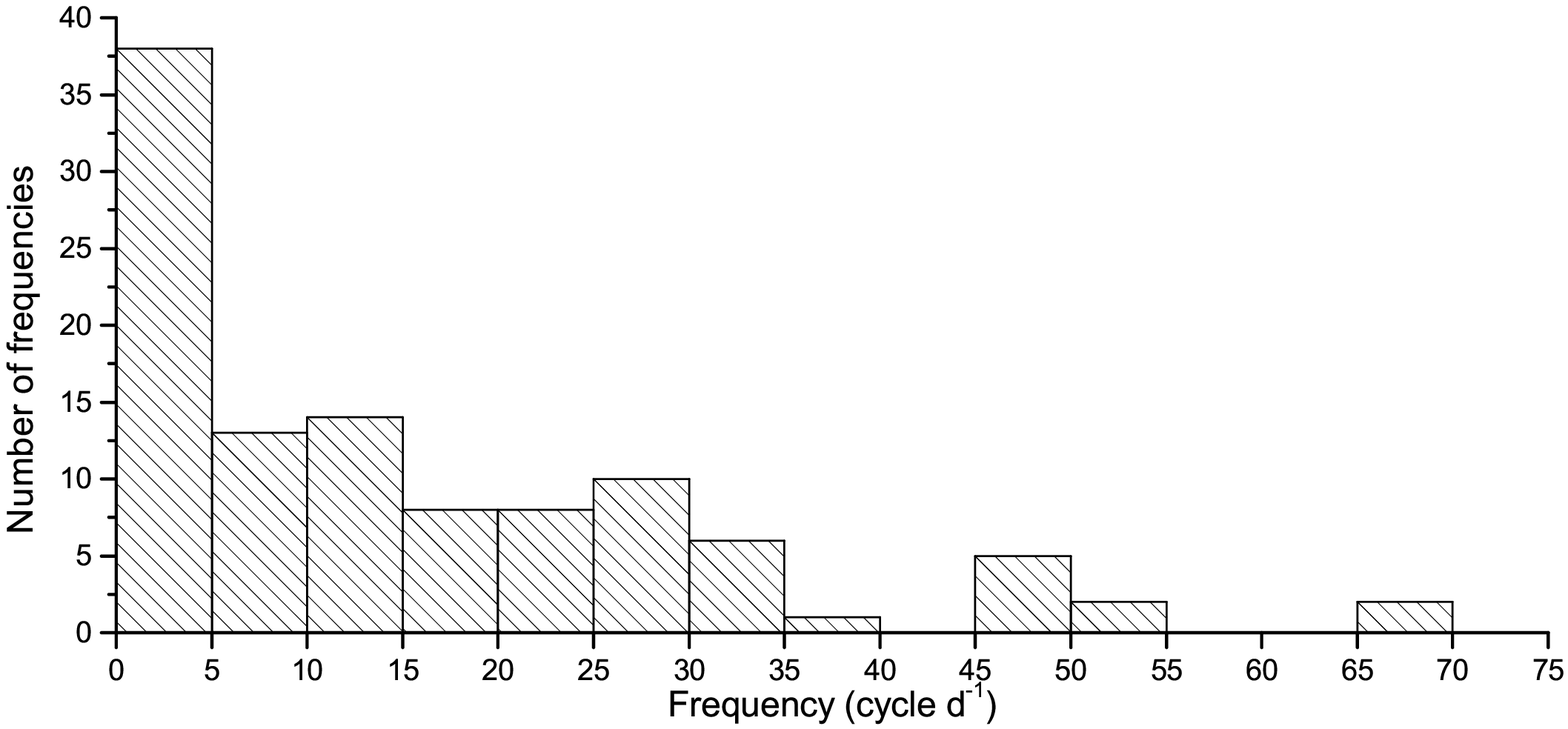}\\
\end{tabular}
\caption{Periodograms (left panels) and frequencies distributions (right panels) for the studied systems. The independent frequencies and the strong frequencies that are connected to the $P_{\rm orb}$ are also indicated.}
\label{fig:FS}
\end{figure*}

\begin{figure*}
\centering
\begin{tabular}{c}
\includegraphics[width=18cm]{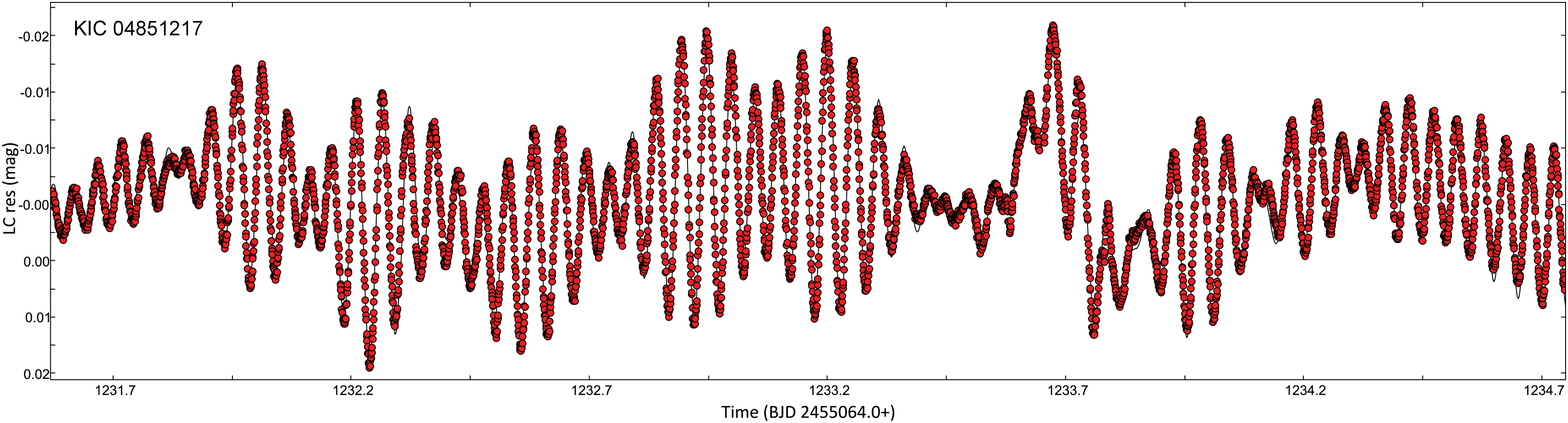}\\
\includegraphics[width=18cm]{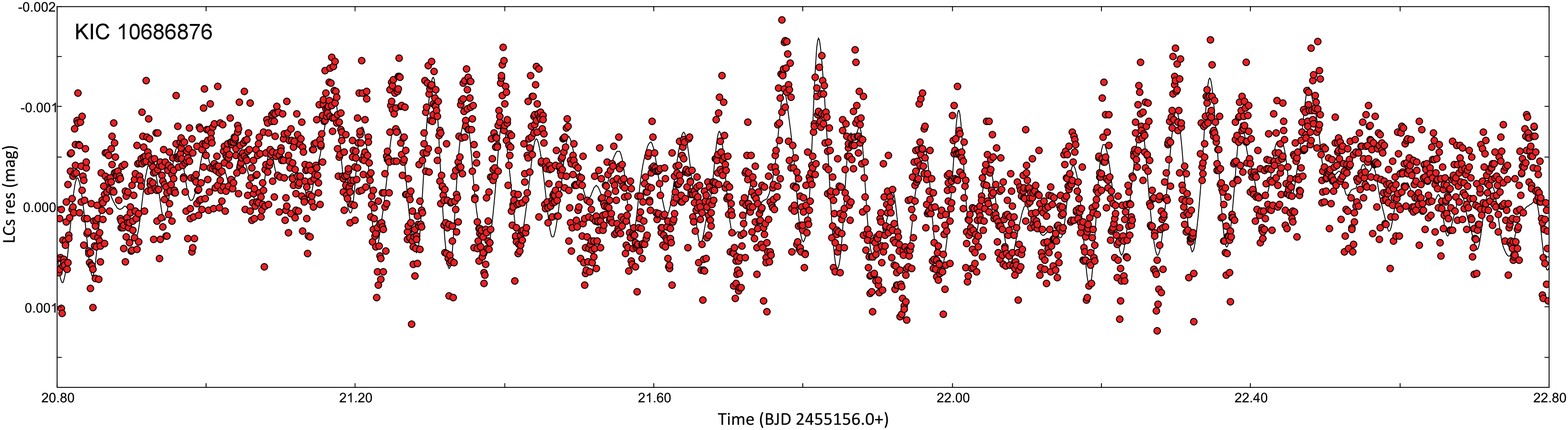}\\
\end{tabular}
\caption{Fourier fitting (solid lines) samples on various data points for both systems.}
\label{fig:FF}
\end{figure*}

The results for the independent frequencies for both systems as well as their mode identification are given also in Table~\ref{tab:IndF}. The frequency values $f_{\rm i}$ with the respective amplitudes $A$, phases $\Phi$, and S/N are listed along with the $Q$, $\rho_{\rm pul}$, $P_{\rm pul}$/$P_{\rm orb}$, $l$-degrees, and modes. Similarly, the parameters of the dependent (combination) frequencies are given in Tables~\ref{tab:DepFreqKIC048}-\ref{tab:DepFreqKIC106} in appendix~\ref{sec:App1}. The periodograms of the systems, on which the independent frequencies and the stronger frequencies that are connected to their $P_{\rm orb}$ are indicated, as well as the distributions of the frequencies are given in Fig.~\ref{fig:FS}. Examples of the Fourier fitting on the observed points for both systems are given in Fig.~\ref{fig:FF}. Comments and information for the individual analysis of each system are given in the following subsections.

\begin{figure}
\includegraphics[width=\columnwidth]{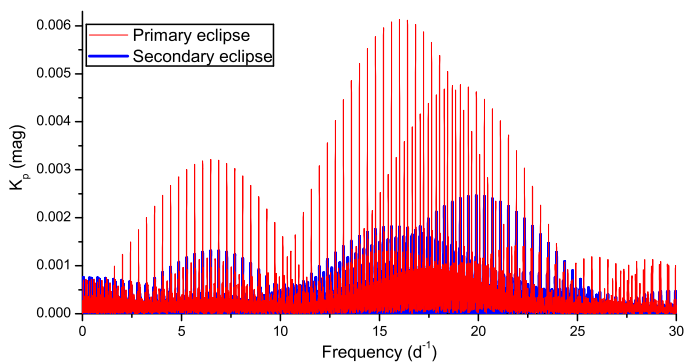}\\
\caption{Periodograms for the data during the eclipses (red=primary; blue=secondary) of KIC~04851217. The amplitudes of the frequencies during the primary eclipse are almost tripled indicating that the pulsating member of the system is the secondary component.}
\label{fig:FS_ecl}
\end{figure}


\begin{table*}
\centering
\caption{Independent oscillation frequencies and mode identification for the pulsating components of both systems. The errors are given in parentheses alongside values and correspond to the last digit(s).}
\label{tab:IndF}
\scalebox{0.97}{
\begin{tabular}{l cccc ccc cc}
\hline																			
$i$	&	  $f_{\rm i}$	&	$A$	&	  $\Phi$	&	S/N	  & 	$Q$	&	$\rho_{\rm pul}$	&	         $P_{\rm pul}$/$P_{\rm orb}\tablefoottext{a}$	&	$l$-degree	&	Pulsation mode	\\
	&	     (d$^{-1}$)	&	(mmag)	&	($\degr$)	&		  & 	(d)     	&	($\rho_{\sun}$)	&		&		&		\\
\hline																			
						 \multicolumn{10}{c}{KIC 04851217}			\\										
\hline																			
1	&	19.092413(1)	&	3.765(7)	&	310.8(1)	&	205.8	  & 	0.018(1)	  & 	0.118(1)	&	0.021	&	0 or 3	&	R or NR $p$	\\
3	&	16.422438(1)	&	2.219(7)	&	350.7(2)	&	140.8	  & 	0.020(1)	  & 		&	0.025	&	1	&	NR $p$	\\
6	&	16.238063(1)	&	1.900(7)	&	100.9(2)	&	125.5	  & 	0.021(1)	  & 		&	0.025	&	1	&	NR $p$	\\
14	&	15.771756(1)	&	0.857(7)	&	31.2(5)	&	63.3	  & 	0.021(1)	  & 		&	0.026	&	0 or 3	&	R or NR $p$	\\
\hline																			
				\multicolumn{10}{c}{KIC 10686876}				\\											
\hline																			
1	&	21.024363(4)	&	0.310(2)	&	328.7(3)	&	129.5	&	0.024(2)	&	0.255(2)	&	0.019	&	2	&	NR $p$	\\
2	&	23.082578(5)	&	0.263(2)	&	5.4(3)	&	106.0	&	0.022(1)	&		&	0.018	&	3	&	NR $p$	\\
3	&	21.456360(9)	&	0.147(2)	&	181.5(6)	&	61.5	&	0.023(2)	&		&	0.019	&	2	&	NR $p$	\\
4	&	19.859294(14)	&	0.092(2)	&	289 (1)	&	39.7	&	0.025(2)	&		&	0.020	&	3	&	NR $p$	\\
7	&	0.184562(18)	&	0.070(2)	&	257(1)	&	8.2	&	2.7(2)	&		&	2.193	&		&		\\
\hline																																																																																																																								\end{tabular}}
\tablefoot{R=radial; NR=non radial; p=pressure mode,
\tablefoottext{a}{Error values are of 10$^{-7}$-10$^{-8}$ order of magnitude.}}
\end{table*}

\subsection{KIC 04851217}
\label{sec:KIC048freqdetails}
The results of the LC modelling (Section~\ref{sec:LCmdl}) for this system show that its components are very similar regarding their absolute properties (i.e. mass, temperature, $\log g$; see Table~\ref{tab:LCmdlAbs}). Therefore, it is not any easy task to understand directly which of the components is the pulsating member, since the physical characteristics of both of them are well inside the range of those of  $\delta$~Scuti stars. Moreover, the system does not exhibit any total eclipses that would help to see when the pulsations are completely vanish. Therefore, as an approximation, the data during the eclipses were used to check the behaviour of the stronger frequencies. In particular, the data points between the orbital phases 0.97-0.03 (primary eclipse) and 0.45-0.51\footnote{Notice that the orbit of the system is slightly eccentric and the secondary minimum occurs at $\Phi_{\rm orb}\sim$0.48.} (secondary eclipse) were used for two different frequency analyses. The periodograms for each data set are plotted together for direct comparison in Fig.~\ref{fig:FS_ecl}. The results show that during the primary eclipse the amplitudes of the frequencies are almost tripled in comparison with the respective ones during the secondary eclipse. During the primary eclipse, the hotter component is occulted, hence, the emitted light comes mainly from the secondary component. Therefore, according to these findings, the most likely oscillating component is the secondary (i.e. the more massive and slightly cooler component).

Four independent frequencies in the regime 15.7-19.1~d$^{-1}$ and another 86 combination frequencies were found. The frequencies are spread in the regions 0.01-6.1~d$^{-1}$ (39), and 15.2-20.3 (41). Two main concentrations of frequencies are distinguished, one between 15-17.5~d$^{-1}$ and another between 18-20.5~d$^{-1}$. Frequencies less than 6.1~d$^{-1}$ are either combinations of others or are multiples of the orbital frequency $f_{\rm orb}$ (e.g. $f_5$, $f_{8}$). According to the theoretical models, the $f_1$ or the $f_{14}$ might be the radial fundamental mode, while the rest frequencies are non-radial pressure modes.

\subsection{KIC 10686876}
\label{sec:KIC106freqdetails}
The pulsating star of this system oscillates in five independent frequencies and in 102 more (depended) ones. All the independent frequencies, except for $f_{7}$, were detected between 19.8-23.1~d$^{-1}$. The $f_{7}$ has a value $\sim0.18$~d$^{-1}$ and no relevance with any other frequency or combination was found. From the total 107 detected frequencies, 38 of them have values less than 5~d$^{-1}$, 59 are uniformly spread in the range 5-35~d$^{-1}$, while ten more lie between 35-68.5~d$^{-1}$. The theoretical models showed that all independent frequencies, except for $f_{7}$, are non-radial $p$-modes.

The distribution of the frequencies plausibly arises the scenario about possible  $\delta$~Scuti-$\gamma$~Doradus hybrid behaviour. According to the criteria of \citet{UYT11}, an hybrid star of this type has to (a) present frequencies in both regimes, (b) the amplitudes in the two domains must be either comparable or the amplitudes should not differ by more than a factor of 5-7, and (c) at least two independent frequencies should be detected in both regions with amplitudes higher than 100~ppm. In the case of KIC~10686876, $f_{7}$ is the only independent frequency in the $\gamma$~Doradus domain and its amplitude is lower by a factor of $\sim4.4$ in comparison with that of $f_1$. Therefore, although the criteria (a) and (b) are satisfied, the (c) is not. Hence, this pulsating star is not likely an hybrid. However, given the large amount of adjusted parameters in the LC modelling (see Section~\ref{sec:KIC106mdldetails}), this frequency may be an artifact. It should be noticed that $f_{\rm orb} \sim 2.1 f_{7}$. All the other low frequencies can be derived from the combinations of the independent ones or are connected to the $f_{\rm orb}$ (i.e. integer multiples). Another possibility for the origin of $f_{7}$ is that it is a $g$-mode oscillation caused by the gravity influence of the secondary component.


\section{Evolution, comparison with similar systems, and correlations}
\label{sec:Evol}

This section includes the evolutionary status of the studied systems within the mass-radius ($M-R$) and Hertzsprung-Russell ($HR$) diagrams and a comparison of their properties with other similar systems. For the latter comparison, the parameters of the pulsating stars and their host systems, as given in \citet[][Tables~1 and 2 therein]{LIAN17}, were used. However, since almost three years have been passed from that publication, new cases of such systems were collected from the literature in order to enlarge the sample. Updated parameters for seven already known cases, 14 new ones, and the systems of the present study are listed in Table~\ref{tab:DEBsUPD}. This table includes also for each system: The name, the orbital period ($P_{\rm orb}$), and the type of variability ($ToV$), while, particularly, for the  $\delta$~Scuti components are given: The dominant pulsation frequency ($f_{\rm dom}$), the mass ($M_{\rm pul}$), the radius ($R_{\rm pul}$), the temperature ($T_{\rm pul}$), and the gravity acceleration value ($\log g_{\rm pul}$).

\begin{figure}
\includegraphics[width=\columnwidth]{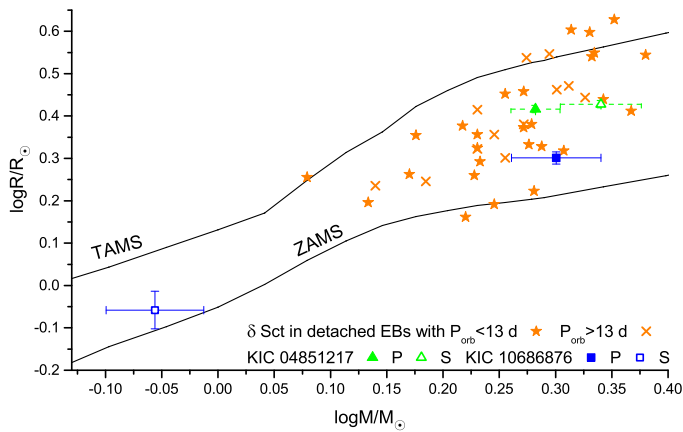}
\caption{Location of the primary (filled symbols) and secondary (empty symbols) components of KIC~04851217 (triangles) and KIC~10686876 (squares) within the mass-radius diagram. The stars and the `x' symbols denote the  $\delta$~Scuti components of other detached systems with $P_{\rm orb}$ shorter and longer than 13~d, respectively  \citep[taken from][and Table~\ref{tab:DEBsUPD}]{LIAN17}, while the black solid lines represent the boundaries of main-sequence.}
\label{fig:MR}
\end{figure}

\begin{figure}
\includegraphics[width=\columnwidth]{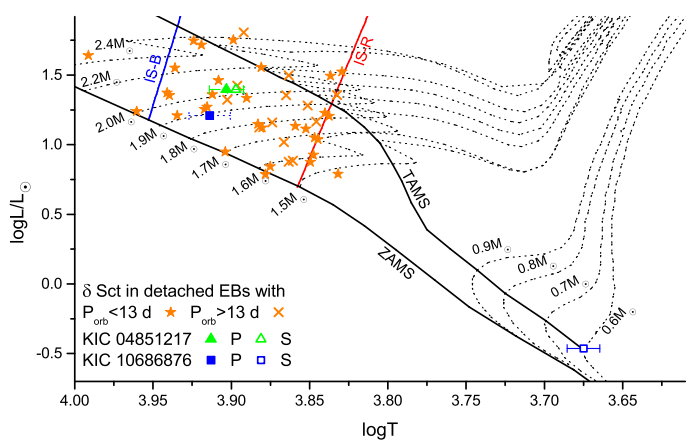}
\caption{Location of the components of both studied systems within the $HR$ diagram. Symbols and black solid lines have the same meaning as in Fig.~\ref{fig:MR}. Dotted lines denote the evolutionary tracks for stars with masses between 0.6-0.9~$M_{\sun}$ and 1.5-2.4~$M_{\sun}$ \citep[taken from][]{GIR00} and the colored solid lines (B=Blue, R=Red) the boundaries of the instability strip \citep[IS; taken from][]{SOY06b}.}
\label{fig:HR}
\end{figure}

The positions of the components of the studied systems along with the  $\delta$~Scuti stars of other detached binaries within the $M-R$ and $HR$ diagrams are given in Figs.~\ref{fig:MR} and \ref{fig:HR}, respectively. All the components are located on the main-sequence. The primaries of both systems and the secondary of KIC~04851217 follow very well the theoretical evolutionary tracks of \citet{GIR00} (see Fig.~\ref{fig:HR}) according to their derived masses and the corresponding error ranges (see Table~\ref{tab:LCmdlAbs}). Therefore, it seems that they have been evolving without any significant interactions so far. The secondary of KIC~10686876, on the other hand, although it was found to lie inside the main-sequence in both evolutionary diagrams, it does not follow very well the evolutionary tracks for $M$=0.9~$M_{\sun}$. It appears closer to zero age main-sequence (ZAMS) in the $M-R$ diagram and closer to the terminal age main-sequence (TAMS) in the $HR$ diagram. However, the errors on its mass, radius, temperature and luminosity values are large enough to accept this discrepancy as possible under(over)estimation of some of its parameters rather than speculating on possible mass transfer from the present primary to the secondary. From evolutionary point of view, the  $\delta$~Scuti stars of the studied systems are well inside the absolute properties ranges of $\delta$~Scuti stars in detached binaries and inside the classical instability strip.


\begin{figure}
\includegraphics[width=\columnwidth]{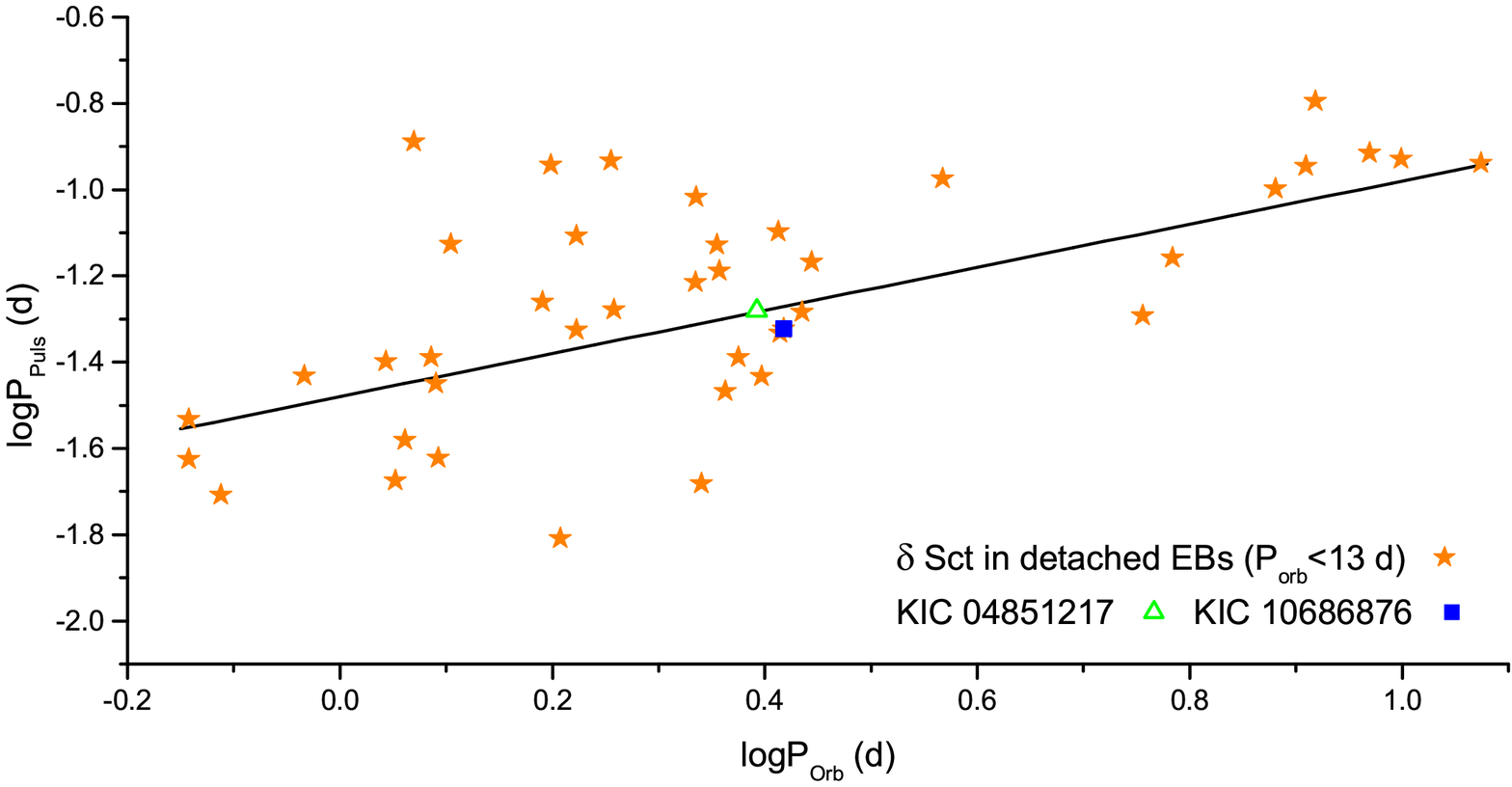}
\caption{Locations of the pulsating components of the studied systems among other  $\delta$~Scuti stars-members of detached systems with $P_{\rm orb}<13$~d within the $P_{\rm orb}-P_{\rm pul}$ diagram. Symbols have the same meaning as in Fig.~\ref{fig:MR}, while the solid line denotes the present linear fit.}
\label{fig:PP}
\includegraphics[width=\columnwidth]{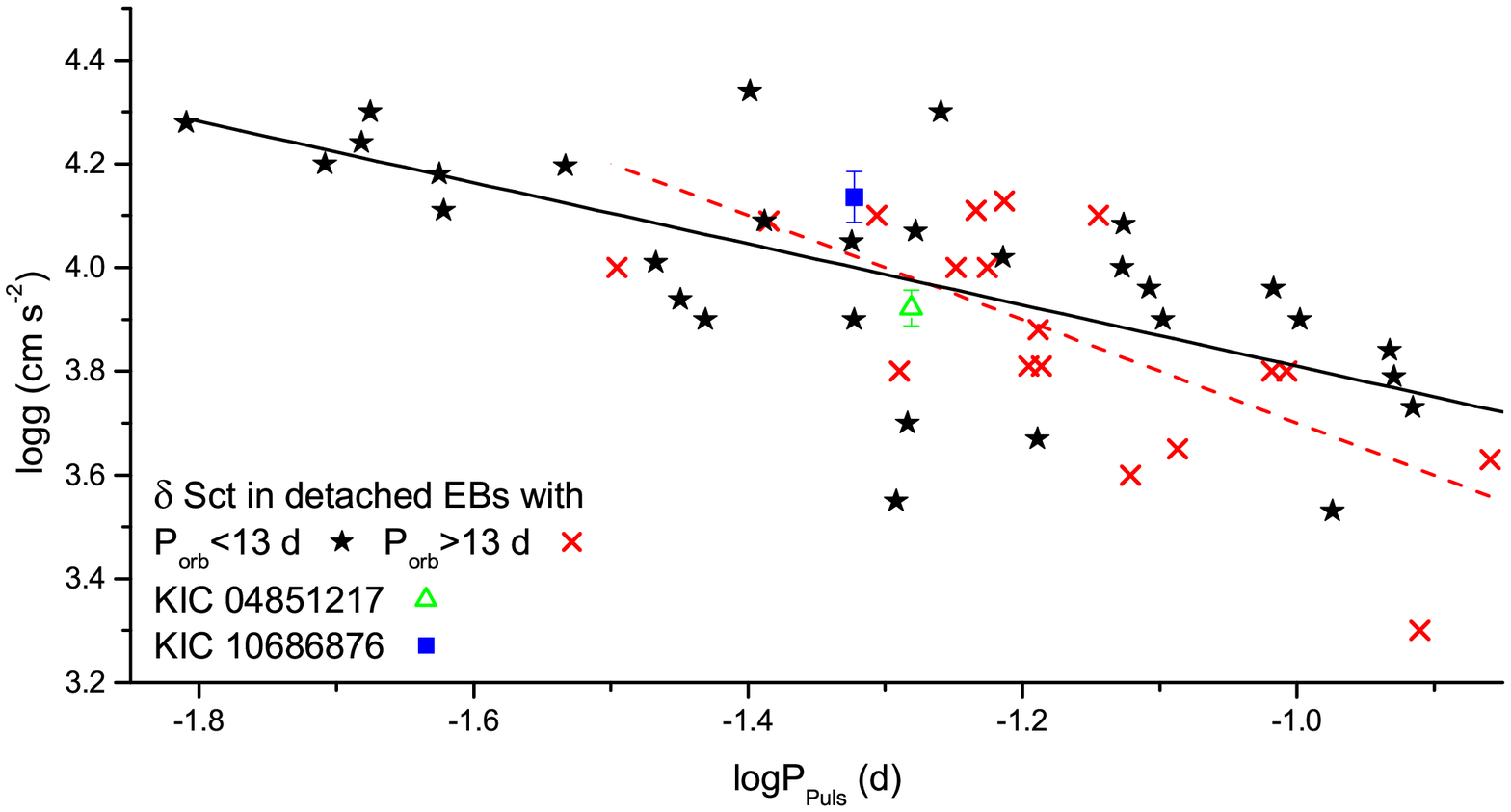}
\caption{Locations of the pulsating components of the studied systems within the $\log g-P_{\rm pul}$ diagram. Symbols have the same meaning as in Fig.~\ref{fig:MR}. Solid and dashed lines denote the present empirical correlations for  $\delta$~Scuti components in detached binary systems with $P_{\rm orb}<13$~d and $P_{\rm orb}>13$~d, respectively.}
\label{fig:gP}
\end{figure}

\begin{table*}
\centering
\caption{New cases of detached EBs with a  $\delta$~Scuti component and known systems with updated properties of their pulsators after the publication of the catalogue of \citet{LIAN17}.}
\label{tab:DEBsUPD}
\begin{tabular}{l cc cc cc cc}
\hline																	
System	&	$P_{\rm orb}$	&	$f_{\rm dom}$	&	$ToV$	&	$M_{\rm pul}$	&	$R_{\rm pul}$	&	$T_{\rm pul}$	&	$\log g_{\rm pul}$	&	Ref.	\\
	&	(d)	&	(d$^{-1}$)	&		&	($M_{\sun}$)	&	($R_{\sun}$)	&	(K)	&	(cm~s$^{-2}$)	&		\\
\hline																	
1SWASP J162842.31+101416.7	&	0.72036	&	42.15	&	SB1+EB	&	1.36	&	1.57	&	7500	&	4.18	&	1	\\
CO Cam	&	1.2709927	&	13.37815	&	SB1+el	&	1.48	&	1.83	&	7070	&	4.08	&	2	\\
TU CMa	&	1.1278	&	47.32	&	SB1+EB	&	1.76	&	1.55	&	8014	&	4.3	&	3, 4	\\
V1224 Cas	&	2.27537	&	15.4516	&	EB	&	2.16	&	3.54	&	8395	&	3.67	&	5	\\
OO Dra	&	1.23838	&	41.867	&	SB2+EB	&	2.03	&	2.08	&	8260	&	4.11	&	6, 7, 8	\\
EPIC 201534540\tablefootmark{a}	&	2.72205	&	19.221	&	SB1+EB	&	2.15	&	3.47	&	7600	&	3.7	&	9	\\
AI Hya	&	8.28994	&	6.2406	&	SB2+EB	&	2.14	&	3.96	&	6864	&	3.57	&	10, 11	\\
KIC 4142768	&	13.9958	&	15.355	&	SB2+EB	&	2.05	&	2.96	&	7327	&	3.81	&	12	\\
KIC 4544587	&	2.1891	&	48.022	&	SB2+EB	&	1.69	&	1.82	&	8600	&	4.24	&	13, 14	\\
KIC 4851217	&	2.47028	&	19.092	&	SB2+EB	&	2.2	&	2.68	&	7890	&	3.92	&	14, 15, pw	\\
KIC 8087799	&	0.9263	&	26.98	&	EB	&	2.2	&	2.75	&	8092	&	3.9	&	16	\\
KIC 8113154	&	2.58688	&	12.521	&	EB	&	1.65	&	2.38	&	7118	&	3.9	&	17	\\
KIC 8197761	&	19.73845	&	17.139	&	SB1+EB	&	1.38	&	1.72	&	7301	&	4.11	&	18, 19	\\
KIC 8262223	&	1.61301	&	64.434	&	SB2+EB	&	1.91	&	1.67	&	9128	&	4.28	&	14, 20	\\
KIC 9285587	&	1.81196	&	18.949	&	SB1+EB	&	1.94	&	2.13	&	8230	&	4.07	&	21	\\
KIC 9592855	&	1.21932	&	24.433	&	SB2+EB	&	1.71	&	1.96	&	7037	&	4.09	&	22	\\
KIC 10661783	&	1.23136	&	28.135	&	SB2+EB	&	2.33	&	2.58	&	7764	&	3.94	&	14, 23, 24	\\
KIC 10686876	&	2.61843	&	21.024	&	SB1+EB	&	2.0	&	2.00	&	8200	&	4.14	&	pw	\\
KIC 10736223	&	1.1050943	&	25.0445	&	SB2+EB	&	1.7	&	1.45	&	7554	&	4.34	&	25	\\
KIC 10989032	&	2.3051	&	29.319	&	SB1+EB	&	2.64	&	2.68	&	8630	&	4.01	&	16	\\
KIC 11401845	&	2.16139	&	16.378	&	EB	&	1.7	&	2.11	&	7590	&	4.02	&	26, 27, 28	\\
EU Peg	&	0.72113	&	34.119	&	EB	&	2.6	&	2.13	&	8730	&	4.2	&	29	\\
TIC 309658221\tablefootmark{b}	&	7.59517	&	9.946	&	EB	&	1.5	&	2.26	&	6993	&	3.9	&	30	\\
\hline																	
\end{tabular}
\tablefoot{SB1/2= Single (1) or double (2) line spectroscopic binary, EB=Eclipsing binary, el=ellipsoidal variable.
\tablefoottext{a}{HD 099458}, \tablefoottext{b}{HD 034954}
\tablebib{(1) \citet{MAX14}, (2) \citet{KUR20}, (3) \citet{GAR17}, (4) \citet{MKR19}, (5) \citet{WAN18}, (6) \citet{LEE18}, (7) \citet{ZHA14}, (8) \citet{DIM5842}, (9) \citet{SKA19}, (10) \citet{EKE14}, (11) \citet{LEE20}, (12) \citet{GUO19}, (13) \citet{HAM13}, (14) \citet{MAT17}, (15) \citet{HEL19}, (16) \citet{ZHA17}, (17) \citet{ZHA19}, (18) \citet{SOW17}, (19) \citet{GAU19}, (20) \citet{GUO17}, (21) \citet{FAI15}, (22) \citet{GUO17b}, (23) \citet{SOU11}, (24) \citet{LEH13},  (25) \citet{CHE20}, (26) \citet{LEE17a}, (27) \citet{GAU14}, (28) \citet{HUB14}, (29) \citet{YAN18}, (30) \citet{LEE19a}, (pw) present work.}}
\end{table*}

The pulsating components of the systems under study were placed among others of similar properties (i.e.  $\delta$~Scuti stars of other detached binaries) in the orbital-pulsational periods diagram ($P_{\rm pul}-P_{\rm orb}$), which is shown in Fig.~\ref{fig:PP}. The sample of systems with $P_{\rm orb}<13$~d (41 in total) for this plot was gathered from the catalogue of \citet{LIAN17} and Table~\ref{tab:DEBsUPD}. A new linear fit between these quantities was made and  resulted in the following relation:
\begin{equation}
\log P_{\rm pul} =-1.45(5) + 0.5(1) \log P_{\rm orb},~{\rm with}~r=0.64,
\end{equation}
where $r$ is the correlation coefficient. The difference of this relation with that of \citet{LIAN17} is negligible with both the slope and the intercept values to vary inside their error ranges. That means that, although the present sample is $58\%$ larger than that of \citet{LIAN17}, the deviation between the fittings is very small, hence, the $P_{\rm pul}-P_{\rm orb}$ correlation can be considered as well established. It should to be noted that both systems under study follow very well the empirical curve (solid line in Fig.~\ref{fig:PP}).

The correlation between the quantities $\log g - \log P_{\rm pul}$ is shown in Fig.~\ref{fig:gP}. However, as a next step to the respective correlation of \citet{LIAN17}, it was found useful to present correlations for two samples based on their $P_{\rm orb}$ values. For this, similarly to the $P_{\rm pul}-P_{\rm orb}$ correlation, the sample of  $\delta$~Scuti stars in detached binaries, for which $\log g$ and $P_{\rm pul}$ values are known, was gathered from the same catalogue and table as before. The whole sample was divided into two subgroups, with the first to contain 33  $\delta$~Scuti stars in detached systems with $P_{\rm orb}<13$~days and the second, 18 cases in systems with longer $P_{\rm orb}$ values. Linear fits on the data points of each group were attempted and produced the following correlations:
\begin{equation}
\log g =3.2(1) - 0.6(1) \log P_{\rm pul},~{\rm with}~r=0.69,
\end{equation}
for systems with $P_{\rm orb}<13$~d, and
\begin{equation}
\log g =2.7(3) - 1.0(3) \log P_{\rm pul},~{\rm with}~r=0.68,
\end{equation}
for systems with $P_{\rm orb}>13$~d.\\
\\
The differences between these two correlations are not negligible. Both the slope and the intercept values have large deviations from each other, indicating that the proximity plays significant role to the pulsations. In particular, the  $\delta$~Scuti stars in systems with $P_{\rm orb}<13$~days have longer $P_{\rm pul}$ values for a given $\log g$. That means that, as the star evolves, the frequency values decrease with a slower rate when the star is in a system with $P_{\rm orb}<13$~days rather than if it was in a system with longer $P_{\rm orb}$. According to the stellar evolution, as a pulsating star evolves through the classical instability strip, its oscillating frequencies values decrease. However, the distribution of the  $\delta$~Scuti stars in detached binaries and their corresponding correlations (Fig.~\ref{fig:gP}) show that the proximity of the companion star is very crucial to the braking of the pulsational frequencies. Therefore, it can be concluded that the close binarity not only initiates the pulsations sooner \citep[i.e. the majority of  $\delta$~Scuti stars in binaries are mostly MS stars;][]{LIAN17}, but also preserves them for a longer period of time.


\section{Summary, discussion and conclusions}
\label{sec:Dis}

In this paper, detailed LCs and frequency analyses for two $Kepler$ detached EBs were presented. The present study increased the current sample of this kind of systems (i.e. $\delta$~Scuti stars in detached binary systems with $P_{\rm orb}<13$~d) by $\sim5\%$. Ground-based spectroscopic observations of these systems allowed to estimate the spectral types of their primaries with an error of one sub-class. Moreover, literature RVs were also used for their modelling. The absolute parameters of the components of both systems as well as their evolutionary stages were also estimated. The pulsational characteristics of their oscillating members were determined and were found to lie well inside the range of properties of the  $\delta$~Scuti stars. For both systems, only a very small fraction of the detected pulsation frequencies (i.e. less than 4.7\%) were identified as independent. This comes in good agreement with the results of \citet{KUR15b} for pulsating stars of different types. Furthermore, a comparison of their absolute properties with theoretical stellar evolutionary models was made, while the oscillation modes of their independent frequencies were also estimated using theoretical models of  $\delta$~Scuti stars.

KIC~04851217 was found to have a slightly eccentric orbit and to contain very similar components in terms of absolute parameters. The components were found to be of A6 spectral class. Both members are on the main-sequence and so far there is no sign of strong interaction (e.g. mass transfer). Due to their similarity, frequency analysis for the data during the eclipses was needed to conclude which one is the pulsator. The results showed that the secondary component is more likely the  $\delta$~Scuti member and pulsates in four independent frequencies (with the dominant one at $\sim$19.09~d$^{-1}$) and in another 86 combination frequencies. Two of the independent frequencies were identified as possible radial modes, while the rest as non-radial pressure modes.

KIC~10686876 was revealed as a most difficult case for analysis, although its $K_1$ was calculated by \citet{MAT17}. The difficulties in analysis concerned the selection of the adjusted parameters since it was listed as a possible flare system \citep{DAV16} and it is candidate for hosting a tertiary component \citep{GIE15, ZAS15, BOR16}. The spectrum of the system indicated that the primary component is of A5 spectral class. Both components were found to lie inside MS. For a better LC fitting, a cool and a hot spots on the surface of the secondary were assumed for the majority of the LCs.

The LCs modeling of KIC~10686876 revealed a third light contribution of $2.9\%$ in slight contrast with the value of $5\%$ determined by \citet{ZAS15}. However, the present results for the hypothesis of a tertiary component can be considered more accurate, since the value of \citet{ZAS15} was based on the assumption of $q=1$. From the present LC solution, and by taking into account the level of light contamination of $1.7\%$ in the system's field of view, it can be roughly estimated that the clear third light contribution is $\sim1.2\%$. In order to estimate the minimum mass of the potential tertiary component, the findings of \citet{ZAS15} regarding the mass function $f(m_3)=0.0207$~$M_{\sun}$ of the third body, the absolute parameters of the components of the EB ($M_1=2~M_{\sun}$ and $M_2=0.88~M_{\sun}$; Table~\ref{tab:LCmdlAbs}), and the following formula \citep{MAY90} were used:
\begin{equation}
f(m_{3})=\frac{(M_{3}~\sin i_{3})^{3}}{(M_{1}+M_{2}+M_{3})^{2}}
\end{equation}
where $M$ are the masses of the components, and $i_{3}$ the inclination of the tertiary body's orbit. For $i_{3}=90\degr$ (i.e. coplanar orbits of the EB and the third body), the minimum mass of the additional component ($M_{3,~min}$) yields a value of $0.6~M_{\sun}$. Following the formalism of \citet{LIA11, LIA13}, that is assuming that this component is a main-sequence star, and by using the \textsc{InPeVEB} software \citep{LIA15}, the luminosity contribution of the tertiary member to the total light is found to be $\sim1.1\%$. This value comes in very well agreement with the observed one ($L_3/L_{\rm Total}\sim1.2\%$). Therefore, it can be concluded that, the most possible scenario is that another main-sequence component with mass slightly greater than $0.6~M_{\sun}$ in coplanar orbit or with mass $0.6~M_{\sun}$ and slightly non-coplanar orbit revolves around the EB with a period of $\sim6.72$~yr \citep{ZAS15} and a semi-major axis of 4.4~AU. Another possible explanation for the cyclic modulation of the system's $P_{\rm orb}$ could be the magnetic braking mechanism \citep{APP92}. According to the criteria of \citet{LAN02} regarding the quadrupole moment variation $\Delta Q$ and by using the \textsc{InPeVEB} software \citep{LIA15}, it was found that the secondary component has a value of $\Delta Q=1.77\times 10^{51}$~g~cm$^2$, which is outside the range set by criteria. Therefore, the third body existence is the most probable orbital period modulator mechanism of this EB.

The primary component of KIC~10686876 was found to oscillate in five independent frequencies, with the dominant to be 21.02~d$^{-1}$. The theoretical models showed that all frequencies are likely $p$-modes. The independent $f_{7}\sim0.18$~d$^{-1}$ and the majority of the remaining 102 combination frequencies, which were detected in the range 0-5~d$^{-1}$, enabled the scenario about possible $\delta$~Scuti-$\gamma$~Doradus hybrid behaviour. However, using the criteria of \citet{UYT11}, it can be concluded that probably the star is not an hybrid, while, due to the complexity of its LC model and the large amount of free parameters, the reliability of $f_{7}$ might be questioned.

The comparison of the locations of the pulsating stars of the studied systems with those of other similar ones in the $\log P_{\rm orb}-\log P_{\rm pul}$ and $\log g - \log P_{\rm pul}$ diagrams showed that they follow well both the trend of the other data points as well as the empirical fitting curves. Moreover, a sample division of the  $\delta$~Scuti stars in detached systems in those with $P_{\rm orb}<13$~d and in those with longer $P_{\rm orb}$ values showed that the evolution of the frequencies depends on the proximity of the companion star. In particular, the empirical fittings on these two subgroups showed that the pulsations frequencies of the  $\delta$~Scuti stars in detached binaries with $P_{\rm orb}<13$~d decrease with a slower rate than those that belong in systems with wider orbits.

Using the relation between of luminosity and $P_{\rm pul}$ for  $\delta$~Scuti stars of \citet{ZIA19}:
\begin{equation}
M_{\rm V} = -2.94(6) \log P_{\rm pul}-1.34(6),
\label{EQ:PL}
\end{equation}
the distance of each system can be calculated using the distance modulus:
\begin{equation}
D = 10^{\frac{m_{\rm V}-M_{\rm V}}{5}+1},
\label{EQ:D}
\end{equation}
where $m_{\rm V}$ and $M_{\rm V}$ are the apparent and absolute magnitudes, respectively. Using the dominant $P_{\rm pul}$ values given in Table~\ref{tab:IndF}, the $M_{\rm V}$ is feasible to be calculated (Eq.~\ref{EQ:PL}), thus, using the catalogued $m_{\rm V}$ \citep[The Tycho-2 catalogue;][]{HOG00}, the distance of each system can be derived from Eq.~\ref{EQ:D}. Results for both systems are given in Table~\ref{tab:Dist}.

\begin{table}
\centering
\caption{Magnitudes and distances of the systems. The errors are given in parentheses alongside values and correspond to the last digit(s).}
\label{tab:Dist}
\begin{tabular}{cccc}
\hline																	
KIC No	&	$m_{\rm V}$	&	$M_{\rm V}$	&	D	\\
	    &	(mag)	   &	(mag)	    &	(pc)	\\
\hline							
04851217	&	11.24	&	2.4(1)	&	$579^{+38}_{-35}$	\\
10686876	&	11.54	&	2.5(1)	&	$628^{+42}_{-39}$	\\
\hline							
\end{tabular}
\end{table}

Regarding any future studies on KIC~10686876, it could be mentioned that RVs measurements of its secondary will certify or modify the absolute parameters values. However, the acquisition of RV2 can be considered as extremely difficult, since its primary dominates the spectra with a factor of more than $94\%$. In general, detailed LCs and pulsation analyses of other $Kepler$ or $TESS$ detached systems with pulsating components is highly recommended, since the sample of these systems is still small but at the same time it is very promising regarding our knowledge for pulsations in binary systems.

\begin{acknowledgements}
The author acknowledges financial support by the European Space Agency (ESA) under the Near Earth object Lunar Impacts and Optical TrAnsients (NELIOTA) programme, contract no. 4000112943. The author wishes to thank Mrs Maria Pizga for proofreading the text and the anonymous referee for the valuable comments. The `Aristarchos' telescope is operated on Helmos Observatory by the Institute for Astronomy, Astrophysics, Space Applications and Remote Sensing of the National Observatory of Athens. This research has made use of NASA's Astrophysics Data System Bibliographic Services, the SIMBAD, the Mikulski Archive for Space Telescopes (MAST), and the $Kepler$ Eclipsing Binary Catalog data bases.
\end{acknowledgements}

%
%


\bibliography{references} 


\begin{appendix}

\section{Spot migration}
\label{sec:App2}
\begin{figure}
\begin{tabular}{cc}
\multicolumn{2}{c}{\includegraphics[width=\columnwidth]{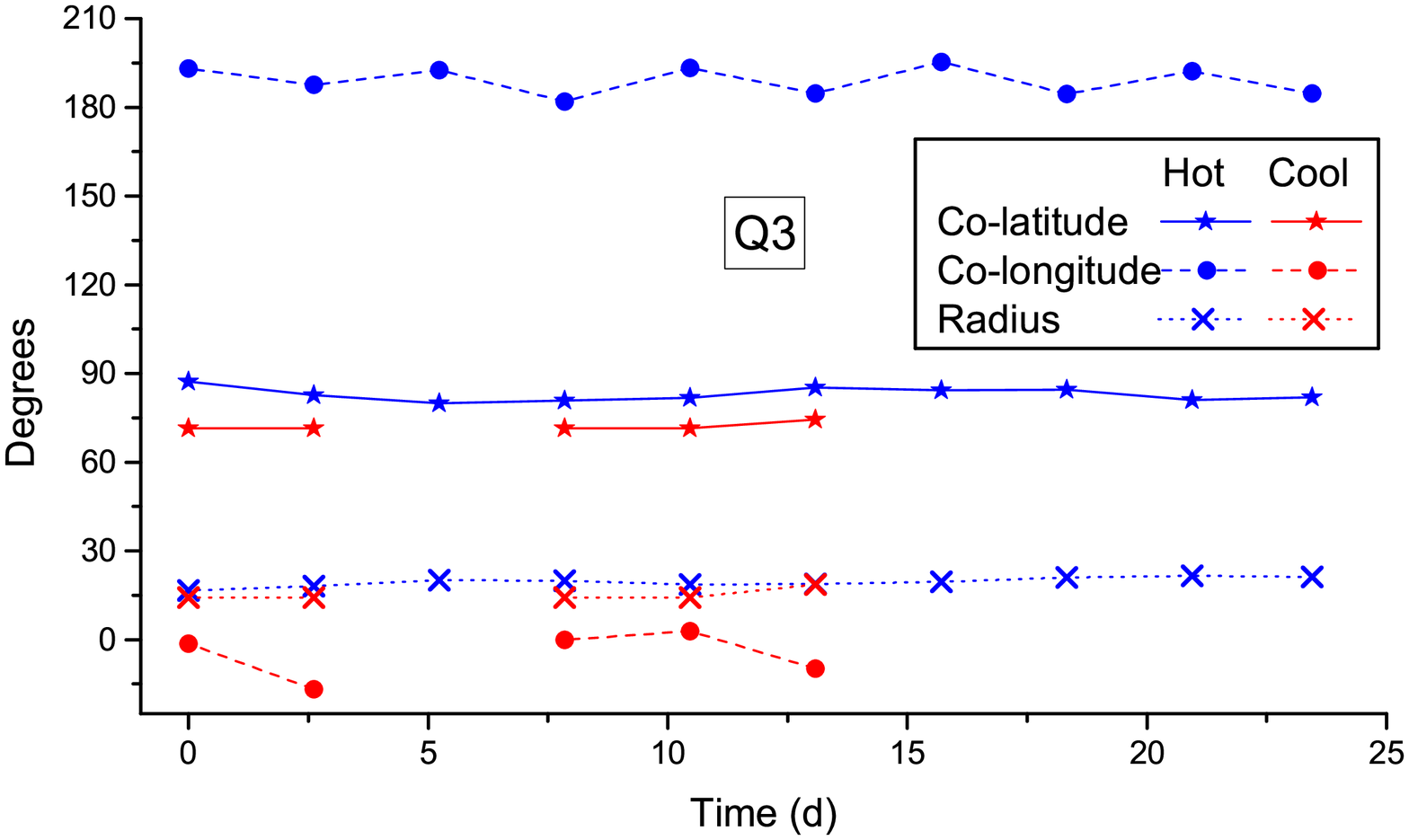}}\\
\multicolumn{2}{c}{\includegraphics[width=\columnwidth]{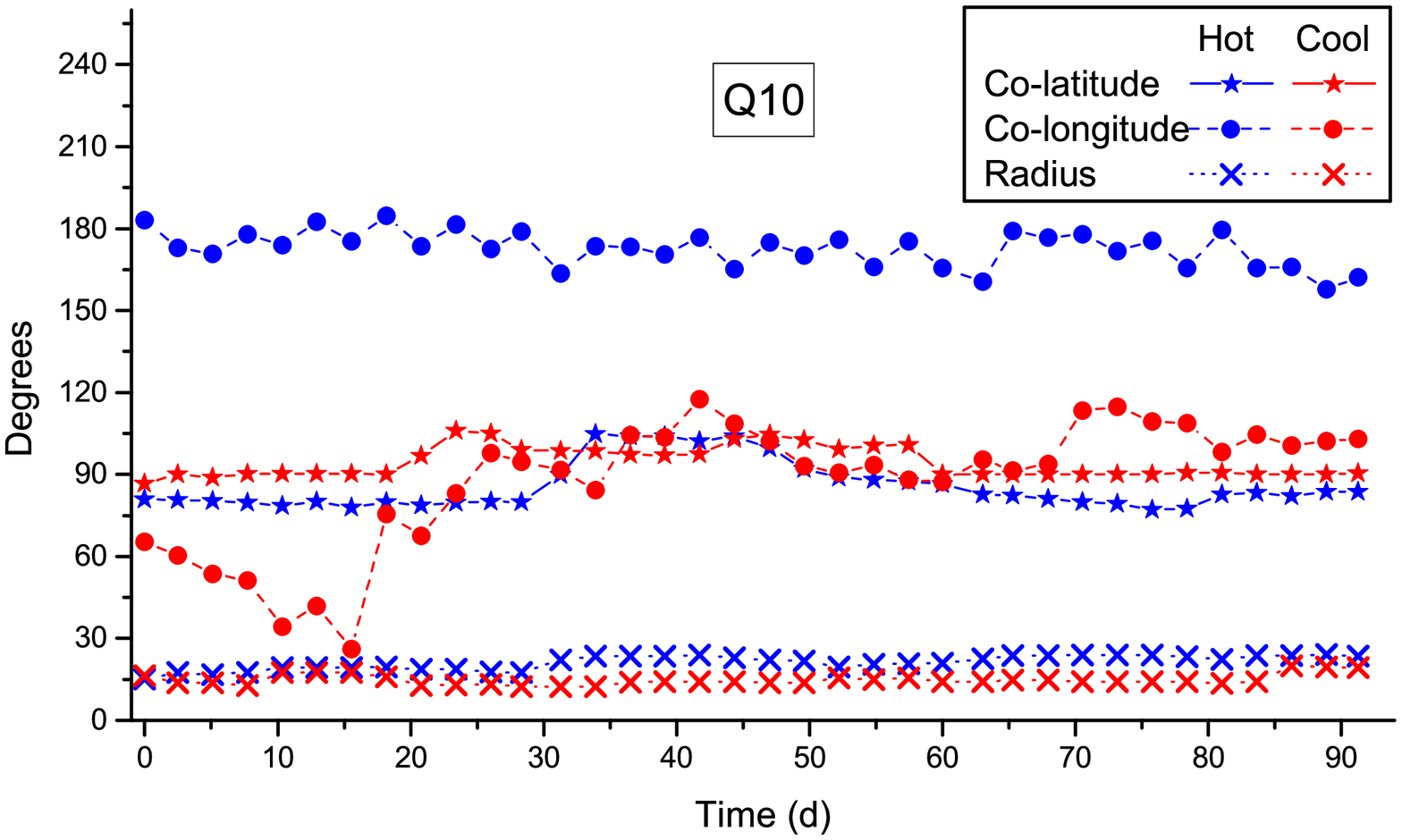}}\\
\includegraphics[width=3cm]{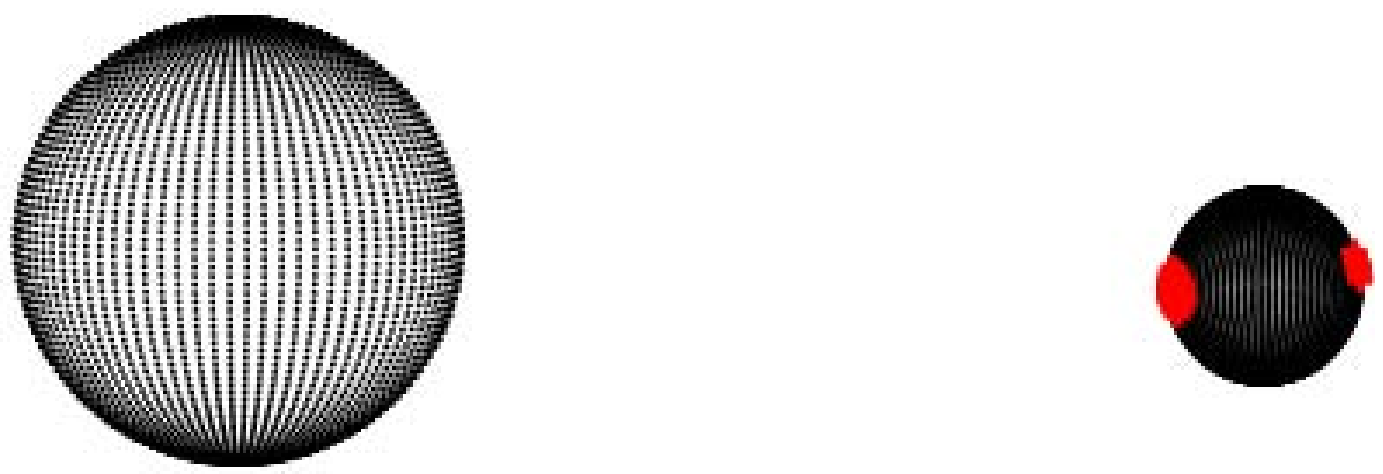}&\hspace{0.5cm}\includegraphics[width=3cm]{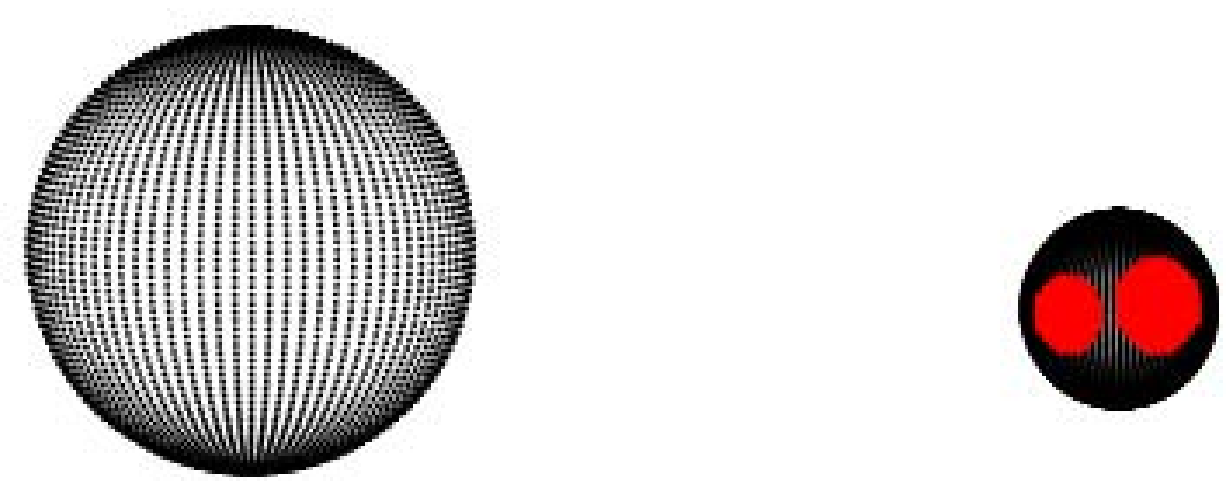}\\
\end{tabular}
\caption{Top panels: Spot migration diagram for KIC~10686876 during Q3 (upper panel) and Q10 (middle panel). Lower left panel shows the positions of both spots on the surface of the secondary at orbital phase 0.15 during the first day of Q10, while lower right panel illustrates the positions of both spots on the same component at an orbital phase of 0.12 during the 91st day of of Q10.}
\label{fig:spotKIC106}
\end{figure}

This appendix includes information for the spots migration in time for KIC~10686876 that was found to present magnetic activity. Details can be found in Section~\ref{sec:LCmdl}. As corresponding timings in Table~\ref{tab:spots} are taken the average BJD values of the points included in the models from which the respective parameters (latitude $Co-lat$, longitude $Co-long$, radius, and temperature factor $Tf$) were calculated. The upper and middle parts of Fig.~\ref{fig:spotKIC106} show the changes of the parameters of all spots over time for each quarter of observations, while the lower part shows the spots on the secondary's surface during different days of observations.

\begin{table*}
\centering
\caption{Spot parameters for KIC~10686876 for the quarters 3 and 10 of $Kepler$ mission. The errors are given in parentheses alongside values and correspond to the last digit(s).}
\label{tab:spots}
\begin{tabular}{c cc cc cc cc}
\hline			
Time	&	Co-lat	&	Long	&	Radius	&	Tf ($\frac{T_{\rm spot}}{T_{\rm eff}}$)	&	Co-lat	&	Long	&	Radius	&	Tf ($\frac{T_{\rm spot}}{T_{\rm eff}}$)	\\
(days)	&	($\degr$)	&	($\degr$)	&	($\degr$)	&		&	($\degr$)	&	($\degr$)	&	($\degr$)	&		\\
\hline
        &\multicolumn{4}{c}{Hot Spot}\          & \multicolumn{4}{c}{Cool Spot}           \\
\hline
\multicolumn{9}{c}{Q3}                    \\
\hline
0.00	&	87(1)	&	193(1)	&	17(1)	&	1.25(1)	&	71(2)	&	359(3)	&	14(2)	&	0.82(4)	\\
2.61	&	83(1)	&	188(2)	&	18(1)	&	1.26(3)	&	71(1)	&	343(2)	&	14(1)	&	0.90(2)	\\
5.23	&	80(1)	&	193(1)	&	20(1)	&	1.25(2)	&		&		&		&		\\
7.85	&	81(1)	&	182(2)	&	20(1)	&	1.26(2)	&	71(1)	&	360(2)	&	14(1)	&	0.93(3)	\\
10.47	&	82(1)	&	193(1)	&	19(1)	&	1.26(2)	&	71(1)	&	3(2)	&	14(1)	&	0.95(2)	\\
13.08	&	85(1)	&	185(1)	&	19(1)	&	1.26(1)	&	74(1)	&	350(2)	&	19(1)	&	0.94(3)	\\
15.71	&	84(1)	&	195(2)	&	20(1)	&	1.24(2)	&		&		&		&		\\
18.33	&	84(1)	&	185(2)	&	21(1)	&	1.22(2)	&		&		&		&		\\
20.95	&	81(1)	&	192(2)	&	22(1)	&	1.22(2)	&		&		&		&		\\
23.45	&	82(1)	&	185(2)	&	21(1)	&	1.22(2)	&		&		&		&		\\
\hline
                                                \multicolumn{9}{c}{Q10}                    \\
\hline
0.00	&	81(1)	&	183(2)	&	15(1)	&	1.32(2)	&	87(2)	&	65(3)	&	16(2)	&	0.71(4)	\\
2.47	&	81(1)	&	173(1)	&	17(1)	&	1.22(2)	&	90(1)	&	60(2)	&	14(1)	&	0.69(2)	\\
5.09	&	80(1)	&	171(1)	&	17(1)	&	1.28(2)	&	89(1)	&	54(2)	&	14(1)	&	0.80(4)	\\
7.71	&	80(6)	&	178(11)	&	18(1)	&	1.24(2)	&	90(1)	&	51(2)	&	13(1)	&	0.89(3)	\\
10.33	&	79(1)	&	174(1)	&	19(1)	&	1.22(2)	&	90(1)	&	34(2)	&	17(1)	&	0.91(2)	\\
12.94	&	80(1)	&	182(1)	&	19(1)	&	1.22(1)	&	90(1)	&	42(2)	&	17(1)	&	0.94(3)	\\
15.56	&	78(1)	&	175(1)	&	19(1)	&	1.23(2)	&	90(1)	&	26(2)	&	17(1)	&	0.94(4)	\\
18.18	&	80(1)	&	185(2)	&	19(1)	&	1.22(2)	&	90(1)	&	76(2)	&	16(1)	&	0.93(2)	\\
20.78	&	79(1)	&	174(2)	&	19(1)	&	1.21(2)	&	97(1)	&	68(1)	&	13(1)	&	0.88(1)	\\
23.43	&	80(1)	&	182(2)	&	19(1)	&	1.23(2)	&	106(2)	&	83(3)	&	13(2)	&	0.87(4)	\\
26.04	&	80(1)	&	172(2)	&	18(1)	&	1.25(2)	&	105(1)	&	98(2)	&	13(1)	&	0.83(2)	\\
28.32	&	80(6)	&	179(11)	&	18(1)	&	1.26(2)	&	99(1)	&	95(2)	&	12(1)	&	0.77(2)	\\
31.28	&	90(1)	&	163(1)	&	22(1)	&	1.18(2)	&	99(1)	&	92(2)	&	12(1)	&	0.87(2)	\\
33.89	&	105(1)	&	174(2)	&	23(1)	&	1.17(2)	&	99(1)	&	84(2)	&	12(1)	&	0.94(2)	\\
36.51	&	104(1)	&	173(1)	&	24(1)	&	1.16(2)	&	97(1)	&	104(2)	&	14(1)	&	0.89(2)	\\
39.11	&	104(2)	&	171(3)	&	24(2)	&	1.17(2)	&	97(1)	&	104(1)	&	14(1)	&	0.88(1)	\\
41.74	&	102(6)	&	177(11)	&	24(1)	&	1.16(2)	&	97(1)	&	118(2)	&	14(1)	&	0.86(2)	\\
44.36	&	104(1)	&	165(1)	&	23(1)	&	1.17(2)	&	103(1)	&	109(1)	&	14(1)	&	0.84(1)	\\
46.99	&	100(6)	&	175(11)	&	22(1)	&	1.14(2)	&	105(2)	&	102(3)	&	14(2)	&	0.83(4)	\\
49.60	&	92(1)	&	170(1)	&	22(1)	&	1.15(1)	&	103(1)	&	93(2)	&	14(1)	&	0.89(2)	\\
52.21	&	89(1)	&	176(1)	&	19(1)	&	1.18(2)	&	99(1)	&	91(2)	&	15(1)	&	0.75(2)	\\
54.84	&	88(1)	&	166(2)	&	20(1)	&	1.18(2)	&	101(1)	&	93(1)	&	15(1)	&	0.72(1)	\\
57.45	&	87(1)	&	175(1)	&	21(1)	&	1.17(1)	&	101(1)	&	88(2)	&	16(1)	&	0.68(2)	\\
60.01	&	86(6)	&	166(11)	&	21(1)	&	1.16(2)	&	90(1)	&	87(2)	&	14(1)	&	0.68(3)	\\
63.04	&	83(1)	&	161(2)	&	22(1)	&	1.17(2)	&	90(1)	&	95(2)	&	14(1)	&	0.88(2)	\\
65.30	&	82(1)	&	179(1)	&	24(1)	&	1.16(1)	&	90(1)	&	91(2)	&	15(1)	&	0.90(4)	\\
67.93	&	81(1)	&	177(2)	&	24(1)	&	1.16(2)	&	90(1)	&	94(2)	&	15(1)	&	0.89(2)	\\
70.54	&	80(1)	&	178(1)	&	24(1)	&	1.15(1)	&	90(3)	&	113(4)	&	14(3)	&	0.88(3)	\\
73.16	&	79(1)	&	172(1)	&	24(1)	&	1.15(2)	&	90(1)	&	115(2)	&	14(1)	&	0.87(2)	\\
75.78	&	77(6)	&	176(11)	&	24(1)	&	1.14(2)	&	90(1)	&	109(2)	&	14(1)	&	0.84(2)	\\
78.39	&	77(1)	&	166(1)	&	23(1)	&	1.15(1)	&	91(6)	&	109(11)	&	14(1)	&	0.80(2)	\\
81.01	&	83(1)	&	179(2)	&	22(1)	&	1.13(2)	&	91(1)	&	98(1)	&	14(1)	&	0.77(1)	\\
83.64	&	83(6)	&	166(11)	&	24(1)	&	1.13(2)	&	90(2)	&	105(3)	&	14(2)	&	0.73(4)	\\
86.26	&	82(1)	&	166(1)	&	24(1)	&	1.13(1)	&	90(1)	&	101(1)	&	20(1)	&	0.77(1)	\\
88.89	&	84(1)	&	158(1)	&	24(1)	&	1.11(2)	&	90(1)	&	102(2)	&	19(1)	&	0.82(3)	\\
91.24	&	84(6)	&	162(11)	&	24(1)	&	1.11(2)	&	90(6)	&	103(11)	&	19(1)	&	0.82(2)	\\
\hline																			
\end{tabular}
\tablefoot{The starting BJDs are 2455157.84891 for Q3 and 55740.99299 for Q10.}
\end{table*}

\section{Lists of combined frequencies}
\label{sec:App1}

Tables~\ref{tab:DepFreqKIC048}-\ref{tab:DepFreqKIC106} contain the values of the depended frequencies $f_{\rm i}$ (where $i$ is an increasing number), semi-amplitudes $A$, phases $\Phi$, and S/N. Moreover, in the last column of these tables, the most possible combination for each frequency is also given.


\begin{table*}
\centering
\caption{Combination frequencies of KIC~04851217. The errors are given in parentheses alongside values and correspond to the last digit(s).}
\label{tab:DepFreqKIC048}
\scalebox{0.90}{
\begin{tabular}{cc cc cl cc cc cl}
\hline																													
$i$	&	$f_{\rm	i}$	&	$A$	&	$\Phi$	&	S/N	&	Combination				&	$i$	&	$f_{\rm	i}$	&	$A$	&	$\Phi$	&	S/N	&	Combination		\\
	&	(d$^{-1}$)	&	(mmag)	&	($\degr$)	&	&						&		&	(d$^{-1}$)	&	(mmag)	&	($\degr$)	&	&				\\
\hline																							
2	&	19.494725(1)	&	2.434(7)	&	303.1(2)	&	134.2	&	$f_1$+$f_{\rm orb}$	&	48	&	19.092389(11)	&	0.278(7)	&	294.4(1.5)	&	15.5	&	$\sim f_1$	\\
4	&	20.304345(1)	&	2.195(7)	&	180.6(2)	&	138.5	&	$f_1$+3$f_{\rm orb}$	&	49	&	16.048157(11)	&	0.268(7)	&	24.1(1.6)	&	18.4	&	$\sim f_{27}$	\\
5	&	0.809564(2)	&	1.960(7)	&	138.9(2)	&	64.9	&	2$f_{\rm orb}$	&	50	&	20.306773(12)	&	0.256(7)	&	250.8(1.7)	&	15.9	&	$\sim f_{4}$	\\
7	&	15.612815(2)	&	1.384(7)	&	292.0(3)	&	106.6	&	$f_3$-$2f_{\rm orb}$	&	51	&	19.090183(12)	&	0.251(7)	&	358.9(1.7)	&	13.8	&	$\sim f_1$	\\
8	&	1.619349(2)	&	1.377(7)	&	351.8(3)	&	97.9	&	4$f_{\rm orb}$	&	52	&	0.026993(12)	&	0.251(7)	&	52.3(1.7)	&	6.4	&	$f_{49}$-$f_{12}$	\\
9	&	2.428987(2)	&	1.259(7)	&	50.0(3)	&	110.0	&	6$f_{\rm orb}$	&	53	&	0.052588(12)	&	0.255(7)	&	98.1(1.7)	&	6.6	&	$\sim f_{44}$	\\
10	&	1.214457(3)	&	1.128(7)	&	183.7(4)	&	52.6	&	3$f_{\rm orb}$	&	54	&	6.072284(12)	&	0.245(7)	&	63.2(1.7)	&	45.4	&	15$f_{\rm orb}$	\\
11	&	3.238625(3)	&	1.021(7)	&	99.0(4)	&	103.0	&	8$f_{\rm orb}$	&	55	&	18.272528(12)	&	0.242(7)	&	109.7(1.8)	&	12.7	&	$f_{40}$+$f_9$	\\
12	&	16.017634(3)	&	0.919(7)	&	126.0(5)	&	64.1	&	$f_3$-$f_{\rm orb}$	&	56	&	20.145257(12)	&	0.240(7)	&	248.2(1.8)	&	15.0	&	$f_{2}$+$f_{3}$-$f_{14}$	\\
13	&	0.404745(3)	&	0.911(7)	&	180.0(5)	&	24.3	&	$f_{\rm orb}$	&	57	&	17.721639(13)	&	0.237(7)	&	80.6(1.8)	&	13.3	&	$f_{4}$+$f_{3}$-$f_{28}$	\\
15	&	17.822035(4)	&	0.828(7)	&	212.8(5)	&	46.9	&	$f_{10}$+2$f_3$-$f_6$	&	58	&	0.139451(13)	&	0.234(7)	&	329.8(1.8)	&	6.0	&	$f_{3}$-$f_{44}$-$f_{6}$	\\
16	&	15.428352(4)	&	0.797(7)	&	259.1(5)	&	65.7	&	$f_6$-2$f_{\rm orb}$	&	59	&	18.046068(13)	&	0.233(7)	&	158.5(1.8)	&	13.3	&	$f_{3}$+$f_8$	\\
17	&	16.827272(4)	&	0.762(7)	&	202.7(6)	&	47.2	&	$f_3$+$f_{\rm orb}$	&	60	&	19.494193(14)	&	0.214(7)	&	177.5(2.0)	&	11.8	&	$\sim f_2$	\\
18	&	4.048263(4)	&	0.745(7)	&	145.7(6)	&	90.6	&	10$f_{\rm orb}$	&	61	&	19.094449(14)	&	0.212(7)	&	254.4(2.0)	&	11.8	&	$\sim f_1$	\\
19	&	19.092904(4)	&	0.689(7)	&	279.6(6)	&	36.9	&	$\sim f_1$	&	62	&	0.034568(15)	&	0.206(7)	&	345.8(2.1)	&	5.3	&	$f_{36}$-$f_{3}$	\\
20	&	2.833806(4)	&	0.670(7)	&	217.5(6)	&	64.7	&	7$f_{\rm orb}$	&	63	&	0.036003(15)	&	0.202(7)	&	289.8(2.1)	&	5.2	&	$\sim f_{62}$	\\
21	&	3.643444(5)	&	0.665(7)	&	265.6(6)	&	70.8	&	9$f_{\rm orb}$	&	64	&	20.023605(15)	&	0.198(7)	&	67.1(2.1)	&	11.6	&	$f_{14}$+$f_{4}$-$f_{27}$	\\
22	&	17.518715(5)	&	0.603(7)	&	2.3(7)	&	33.2	&	$f_{5}$+2$f_6$-$f_{14}$	&	65	&	16.453013(15)	&	0.197(7)	&	201.2(2.1)	&	12.5	&	$\sim f_{36}$	\\
23	&	2.024168(5)	&	0.597(7)	&	165.2(7)	&	49.3	&	5$f_{\rm orb}$	&	66	&	0.210904(15)	&	0.193(7)	&	320.8(2.2)	&	5.0	&	$f_{65}$-$f_{6}$	\\
24	&	4.453082(5)	&	0.570(7)	&	313.1(7)	&	79.6	&	11$f_{\rm orb}$	&	67	&	0.237861(16)	&	0.187(7)	&	107.2(2.3)	&	4.9	&	$f_{1}$-$f_{42}$	\\
25	&	19.578481(5)	&	0.566(7)	&	280.7(7)	&	32.2	&	2$f_{1}$-$f_{14}$-$f_{20}$	&	68	&	18.621376(16)	&	0.187(7)	&	131.5(2.3)	&	9.7	&	$f_{42}$-$f_{67}$	\\
26	&	4.857827(6)	&	0.490(7)	&	211.4(9)	&	77.6	&	12$f_{\rm orb}$	&	69	&	20.303831(16)	&	0.184(7)	&	59.1(2.3)	&	11.4	&	$\sim f_{4}$	\\
27	&	16.051835(6)	&	0.479(7)	&	226.3(9)	&	32.6	&	2$f_6$-$f_3$	&	70	&	0.282395(16)	&	0.184(7)	&	300.7(2.3)	&	4.8	&	2$f_{58}$	\\
28	&	19.007954(7)	&	0.443(7)	&	61.5(1.0)	&	23.5	&	$f_{11}$+$f_{14}$	&	71	&	0.255807(16)	&	0.182(7)	&	343.1(2.3)	&	4.8	&	$f_{44}$+$f_{66}$	\\
29	&	19.137990(7)	&	0.427(7)	&	283.3(1.0)	&	23.1	&	$f_{22}$+$f_{8}$	&	72	&	1.213942(17)	&	0.180(7)	&	55.1(2.4)	&	8.3	&	$\sim f_{10}$	\\
30	&	5.262646(7)	&	0.414(7)	&	24.4(1.0)	&	77.0	&	13$f_{\rm orb}$	&	73	&	0.112237(17)	&	0.179(7)	&	294.3(2.4)	&	4.6	&	$f_{58}$-$f_{52}$	\\
31	&	18.328353(8)	&	0.391(7)	&	57.9(1.1)	&	21.2	&	$f_{22}$+$f_5$	&	74	&	0.017726(16)	&	0.185(7)	&	264.0(2.3)	&	4.8	&	$f_{44}$-$f_{52}$	\\
32	&	17.047627(8)	&	0.389(7)	&	5.5(1.1)	&	22.7	&	$f_6$+2$f_{\rm orb}$	&	75	&	0.096571(17)	&	0.177(7)	&	189.7(2.4)	&	4.6	&	2$f_{44}$	\\
33	&	19.423732(8)	&	0.383(7)	&	179.2(1.1)	&	20.4	&	$f_{25}$+$f_{7}$-$f_{14}$	&	76	&	19.082239(17)	&	0.177(7)	&	320.1(2.4)	&	9.8	&	$f_{5}$+$f_{55}$	\\
34	&	15.242270(8)	&	0.374(7)	&	134.4(1.1)	&	32.0	&	$f_{27}$-$f_5$	&	77	&	17.232164(17)	&	0.175(7)	&	15.2(2.4)	&	10.0	&	$f_{3}$+$f_{5}$	\\
35	&	16.642808(8)	&	0.353(7)	&	311.6(1.2)	&	22.0	&	$f_6$+$f_{\rm orb}$	&	78	&	0.002648(17)	&	0.175(7)	&	181.5(2.4)	&	4.5	&	$f_{1}$-$f_{19}$	\\
36	&	16.456727(9)	&	0.349(7)	&	85.6(1.2)	&	22.5	&	$f_{13}$+$f_{27}$	&	79	&	0.104588(17)	&	0.175(7)	&	37.4(2.4)	&	4.5	&	2$f_{53}$	\\
37	&	16.861472(9)	&	0.346(7)	&	80.6(1.2)	&	21.5	&	$f_{5}$+$f_{27}$	&	80	&	0.042806(17)	&	0.177(7)	&	141.2(2.4)	&	4.5	&	$f_{29}$-$f_{1}$	\\
38	&	19.817592(9)	&	0.343(7)	&	288.9(1.2)	&	18.8	&	$f_{5}$+$f_{28}$	&	81	&	0.064798(17)	&	0.176(7)	&	268.1(2.4)	&	4.5	&	2$f_{62}$	\\
39	&	18.768843(9)	&	0.340(7)	&	223.0(1.2)	&	17.5	&	$f_{25}$-$f_{5}$	&	82	&	20.046773(18)	&	0.166(7)	&	245.3(2.5)	&	9.6	&	$f_{4}$-$f_{71}$	\\
40	&	15.842364(10)	&	0.310(7)	&	326.8(1.4)	&	22.7	&	$f_{15}$+$f_{22}$-$f_2$	&	83	&	0.093556(19)	&	0.161(7)	&	25.6(2.6)	&	4.1	&	$\sim f_{75}$	\\
41	&	18.687644(10)	&	0.298(7)	&	99.9(1.4)	&	15.4	&	$f_1$-$f_{\rm orb}$	&	84	&	0.264412(19)	&	0.160(7)	&	69.7(2.6)	&	4.2	&	$f_{13}$-$f_{58}$	\\
42	&	18.855706(10)	&	0.297(7)	&	214.3(1.4)	&	15.4	&	$f_{3}$+$f_9$	&	85	&	2.429649(19)	&	0.158(7)	&	126.2(2.7)	&	14.1	&	$\sim f_{9}$	\\
43	&	15.833171(10)	&	0.296(7)	&	106.2(1.4)	&	21.3	&	$f_6$-$f_{\rm orb}$	&	86	&	0.009047(19)	&	0.158(7)	&	136.0(2.7)	&	4.0	&	2$ f_{78}$	\\
44	&	0.049573(10)	&	0.289(7)	&	167.2(1.5)	&	7.5	    &	$f_{29}$-$f_1$	&	87	&	0.121431(19)	&	0.158(7)	&	113.5(2.7)	&	4.1	&	$f_{13}$-$f_{70}$	\\
45	&	16.176502(11)	&	0.284(7)	&	246.9(1.5)	&	19.9	&	$f_{14}$+$f_{\rm orb}$	&	88	&	19.214041(19)	&	0.157(7)	&	344.7(2.7)	&	8.5	&	$f_{1}$+$f_{87}$	\\
46	&	5.667391(11)	&	0.282(7)	&	273.1(1.5)	&	53.7	&	14$f_{\rm orb}$	&	89	&	16.238504(19)	&	0.156(7)	&	121.9(2.7)	&	10.8	&	$\sim f_{6}$	\\
47	&	16.235930(11)	&	0.281(7)	&	167.7(1.5)	&	18.6	&	$\sim f_6$	&	90	&	0.221091(20)	&	0.153(7)	&	28.4(2.8)	&	4.0	&	2$f_{73}$	\\
\hline																							
\end{tabular}}																							
\end{table*}																							
																							

\begin{table*}
\centering
\caption{Combination frequencies of KIC~10686876. The errors are given in parentheses alongside values and correspond to the last digit(s).}
\label{tab:DepFreqKIC106}
\scalebox{0.90}{
\begin{tabular}{cc cc cl cc cc cl}
\hline																													
$i$	&	$f_{\rm	i}$	&	$A$	&	$\Phi$	&	S/N	&	Combination				&	$i$	&	$f_{\rm	i}$	&	$A$	&	$\Phi$	&	S/N	&	Combination		\\
	&	(d$^{-1}$)	&	(mmag)	&	($\degr$)	&	&						&		&	(d$^{-1}$)	&	(mmag)	&	($\degr$)	&	&				\\
\hline
5	&	0.76454(1)	&	0.090(2)	&	190(1)	&	10.8	&	$2f_{\rm orb}$	&	57	&	28.64387(5)	&	0.024(2)	&	229(4)	&	10.1	&	$f_{36}$+$f_{50}$	\\
6	&	1.14637(2)	&	0.075(2)	&	306(1)	&	9.4	&	$3f_{\rm orb}$	&	58	&	29.40610(5)	&	0.025(2)	&	350(4)	&	10.8	&	$f_{5}$+$f_{57}$	\\
8	&	6.87429(2)	&	0.069(2)	&	80(1)	&	25.5	&	$18f_{\rm orb}$	&	59	&	26.35230(5)	&	0.024(2)	&	265(4)	&	9.2	&	$f_{33}$+$f_{4}$	\\
9	&	0.18707(2)	&	0.069(2)	&	239(1)	&	8	&	$\sim f_7$	&	60	&	37.42558(5)	&	0.024(2)	&	293(4)	&	12.9	&	$f_{56}$+$f_{13}$	\\
10	&	3.81787(2)	&	0.065(2)	&	229(1)	&	17.8	&	$10f_{\rm orb}$	&	61	&	25.59013(5)	&	0.023(2)	&	128(4)	&	8.9	&	$f_{23}$+$f_{4}$	\\
11	&	13.74711(2)	&	0.061(2)	&	240(1)	&	23.5	&	2$f_8$	&	62	&	27.88166(5)	&	0.023(2)	&	109(4)	&	9.7	&	$f_{57}$-$f_{5}$	\\
12	&	5.34682(2)	&	0.055(2)	&	162(2)	&	17.7	&	$14f_{\rm orb}$	&	63	&	13.36507(6)	&	0.023(2)	&	196(4)	&	8.8	&	$f_{4}$-$f_{33}$	\\
13	&	19.09526(2)	&	0.055(2)	&	26(2)	&	24.4	&	$f_4$-$f_5$	&	64	&	22.53585(6)	&	0.022(2)	&	320(4)	&	8.9	&	$f_{58}$-$f_{8}$	\\
14	&	15.27473(2)	&	0.054(2)	&	123(2)	&	22.6	&	$f_{13}$-$f_{10}$	&	65	&	30.16827(6)	&	0.022(2)	&	119(4)	&	9.4	&	$f_{58}$+$f_{5}$	\\
15	&	0.95177(2)	&	0.051(2)	&	165(2)	&	6.1	&	$f_5$+$f_7$	&	66	&	8.02002(6)	&	0.022(2)	&	263(4)	&	8.4	&	$f_{6}$+$f_{8}$	\\
16	&	31.35894(3)	&	0.050(2)	&	218(2)	&	21.3	&	2$f_1$-$f_{10}$-$f_{8}$	&	67	&	3.06137(6)	&	0.021(2)	&	223(4)	&	4.9	&	$f_{8}$-$f_{10}$	\\
17	&	0.58989(3)	&	0.050(2)	&	79(2)	&	5.7	&	$f_{10}$+$f_{4}$-$f_{2}$	&	68	&	5.72671(6)	&	0.020(2)	&	317(5)	&	6.6	&	$\sim f_{23}$	\\
18	&	0.39424(3)	&	0.048(2)	&	251(2)	&	5.6	&	$f_5$-2$f_7$	&	69	&	1.52149(7)	&	0.019(2)	&	196(5)	&	4.6	&	$\sim f_{27}$	\\
19	&	12.22112(3)	&	0.047(2)	&	358(2)	&	18.4	&	$f_{12}$+$f_{8}$	&	70	&	11.07033(7)	&	0.020(2)	&	228(5)	&	7.6	&	$f_{19}$-$f_{6}$	\\
20	&	0.55797(3)	&	0.047(2)	&	203(2)	&	5.5	&	$f_6$-$f_{17}$	&	71	&	24.44365(7)	&	0.019(2)	&	300(5)	&	7.5	&	$2f_{19}$	\\
21	&	3.05406(3)	&	0.047(2)	&	110(2)	&	10.5	&	$8f_{\rm orb}$	&	72	&	11.84059(7)	&	0.019(2)	&	325(5)	&	7.3	&	$f_{34}$+$f_{6}$	\\
22	&	0.95576(3)	&	0.046(2)	&	145(2)	&	5.7	&	$\sim f_{15}$	&	73	&	3.04838(7)	&	0.019(2)	&	341(5)	&	4.3	&	$2f_{27}$	\\
23	&	5.72703(3)	&	0.046(2)	&	264(2)	&	15.4	&	$15f_{\rm orb}$	&	74	&	16.03487(7)	&	0.019(2)	&	341(5)	&	8.1	&	$f_{14}$+$f_{5}$	\\
24	&	8.40029(3)	&	0.046(2)	&	327(2)	&	18.3	&	$22f_{\rm orb}$	&	75	&	10.30812(7)	&	0.019(2)	&	103(5)	&	7.6	&	$f_{10}$+$f_{33}$	\\
25	&	0.17540(3)	&	0.045(2)	&	300(2)	&	5.2	&	$f_5$-$f_{17}$	&	76	&	19.85737(7)	&	0.019(2)	&	240(5)	&	8.3	&	$\sim f_{4}$	\\
26	&	4.58183(3)	&	0.045(2)	&	303(2)	&	13.1	&	$12f_{\rm orb}$	&	77	&	2.28578(7)	&	0.019(2)	&	326(5)	&	4.1	&	$2f_{28}$	\\
27	&	1.52334(3)	&	0.045(2)	&	250(2)	&	5.9	&	$4f_{\rm orb}$	&	78	&	1.53295(7)	&	0.019(2)	&	139(5)	&	4.6	&	$2f_{5}$	\\
28	&	1.14062(3)	&	0.046(2)	&	329(2)	&	5.8	&	$3f_{\rm orb}$	&	79	&	23.67332(7)	&	0.018(2)	&	212(5)	&	7	&	$f_{17}$+$f_{2}$	\\
29	&	16.80221(3)	&	0.045(2)	&	27(2)	&	18.8	&	2$f_{24}$	&	80	&	22.91106(7)	&	0.018(2)	&	98(5)	&	7.4	&	$f_{2}$-$f_{25}$	\\
30	&	2.29128(3)	&	0.044(2)	&	108(2)	&	7.7	&	$6f_{\rm orb}$	&	81	&	17.56756(7)	&	0.018(2)	&	153(5)	&	7.9	&	$f_{4}$-$f_{30}$	\\
31	&	4.19911(3)	&	0.044(2)	&	74(2)	&	12.4	&	$11f_{\rm orb}$	&	82	&	12.98395(7)	&	0.017(2)	&	313(5)	&	6.7	&	$2f_{33}$	\\
32	&	4.96307(3)	&	0.046(2)	&	160(2)	&	14.1	&	$13f_{\rm orb}$	&	83	&	11.45629(7)	&	0.018(2)	&	104(5)	&	6.9	&	$2f_{23}$	\\
33	&	6.49069(3)	&	0.043(2)	&	45(2)	&	15.5	&	$17f_{\rm orb}$	&	84	&	12.22333(7)	&	0.017(2)	&	202(5)	&	6.6	&	$\sim f_{19}$	\\
34	&	10.69364(3)	&	0.043(2)	&	85(2)	&	16.8	&	2$f_{12}$	&	85	&	4.96264(7)	&	0.018(2)	&	274(5)	&	5.5	&	$\sim f_{32}$	\\
35	&	0.20909(3)	&	0.041(2)	&	119(2)	&	4.9	&	$f_{5}$-$f_{20}$	&	86	&	45.82812(8)	&	0.016(2)	&	17(5)	&	8.3	&	$f_{24}$+$f_{60}$	\\
36	&	3.43545(3)	&	0.040(2)	&	284(2)	&	9.9	&	$9f_{\rm orb}$	&	87	&	24.82776(8)	&	0.016(2)	&	43(5)	&	6.3	&	$f_{61}$-$f_{5}$	\\
37	&	1.90709(3)	&	0.039(2)	&	211(2)	&	5.9	&	$5f_{\rm orb}$	&	88	&	27.11942(8)	&	0.016(2)	&	354(5)	&	6.3	&	$f_{59}$+$f_{5}$	\\
38	&	0.97187(3)	&	0.039(2)	&	52(2)	&	4.7	&	$f_{35}$+$f_5$	&	89	&	1.70332(8)	&	0.016(2)	&	194(6)	&	4	&	$f_{20}$+$f_{6}$	\\
39	&	6.10960(3)	&	0.039(2)	&	153(2)	&	13.6	&	$16f_{\rm orb}$	&	90	&	14.12397(8)	&	0.016(2)	&	347(6)	&	6	&	$f_{14}$-$f_{6}$	\\
40	&	7.25465(3)	&	0.038(2)	&	139(2)	&	14.3	&	$19f_{\rm orb}$	&	91	&	13.75170(8)	&	0.016(2)	&	286(6)	&	5.8	&	$\sim f_{11}$	\\
41	&	5.34446(3)	&	0.036(2)	&	346(2)	&	11.9	&	$\sim f_{12}$	&	92	&	1.70088(8)	&	0.016(2)	&	160(6)	&	4.1	&	$\sim f_{89}$	\\
42	&	0.66023(3)	&	0.036(2)	&	167(2)	&	4.2	&	$f_{17}$+$f_3$-2$f_{34}$	&	93	&	31.70065(8)	&	0.016(2)	&	351(6)	&	5.6	&	$f_{4}$+$f_{72}$	\\
43	&	0.93759(4)	&	0.033(2)	&	311(3)	&	4.1	&	$f_{25}$+$f_5$	&	94	&	27.49797(8)	&	0.016(2)	&	69(6)	&	5.6	&	$2f_{11}$	\\
44	&	14.51077(4)	&	0.032(2)	&	23(3)	&	12.7	&	2$f_{40}$	&	95	&	32.46300(8)	&	0.016(2)	&	89(6)	&	5.6	&	$f_{4}$+$f_{55}$	\\
45	&	0.79410(4)	&	0.032(2)	&	96(3)	&	4.9	&	$f_{17}$+$f_{35}$	&	96	&	30.93053(8)	&	0.015(2)	&	225(6)	&	5.5	&	$f_{5}$+$f_{65}$	\\
46	&	0.57053(4)	&	0.032(2)	&	179(3)	&	4.8	&	$f_{27}$-$f_{15}$	&	97	&	30.55186(8)	&	0.016(2)	&	153(6)	&	5.5	&	$2f_{14}$	\\
47	&	2.66986(4)	&	0.032(2)	&	208(3)	&	6.4	&	$7f_{\rm orb}$	&	98	&	28.25704(8)	&	0.016(2)	&	206(6)	&	5.5	&	$f_{4}$+$f_{24}$	\\
48	&	3.81876(4)	&	0.033(2)	&	29(3)	&	8.8	&	$\sim f_{10}$	&	99	&	53.08525(8)	&	0.015(2)	&	240(6)	&	4.9	&	$f_{40}$+$f_{86}$	\\
49	&	0.72258(4)	&	0.032(2)	&	91(3)	&	3.9	&	$f_{19}$+$f_4$-$f_{16}$	&	100	&	53.84740(8)	&	0.016(2)	&	18(6)	&	5	&	$f_{5}$+$f_{99}$	\\
50	&	25.20590(6)	&	0.022(2)	&	59(4)	&	8.5	&	$\sim f_{39}$	&	101	&	47.73634(8)	&	0.015(2)	&	90(6)	&	4.6	&	$f_{4}$+$f_{62}$	\\
51	&	4.58118(4)	&	0.029(2)	&	115(3)	&	8.6	&	$f_{26}$	&	102	&	46.97408(8)	&	0.016(2)	&	331(6)	&	4.4	&	$f_{4}$+$f_{88}$	\\
52	&	7.63545(5)	&	0.026(2)	&	71(3)	&	10	&	$2f_{10}$	&	103	&	46.20948(8)	&	0.016(2)	&	30(5)	&	4.4	&	$f_{4}$+$f_{59}$	\\
53	&	9.16940(5)	&	0.026(2)	&	178(3)	&	9.9	&	$2f_{26}$	&	104	&	68.36122(8)	&	0.015(2)	&	359(6)	&	4.1	&	$f_{14}$+$f_{99}$	\\
54	&	6.87026(5)	&	0.025(2)	&	278(4)	&	9.4	&	$\sim f_{8}$	&	105	&	29.79306(8)	&	0.015(2)	&	1(6)	&	4.3	&	$f_{34}$+$f_{13}$	\\
55	&	12.60291(5)	&	0.025(2)	&	76(4)	&	9.7	&	$f_{11}$-$f_{6}$	&	106	&	45.44695(8)	&	0.015(2)	&	334(6)	&	4.4	&	$f_{4}$+$f_{61}$	\\
56	&	18.32967(5)	&	0.025(2)	&	308(4)	&	10.7	&	$f_{13}$-$f_{5}$	&	107	&	66.83382(9)	&	0.015(2)	&	63(6)	&	4	&	$f_{4}$+$f_{102}$	\\
\hline																													
\end{tabular}}																													
\end{table*}	
																											
\end{appendix}

\end{document}